 \documentclass[final,5p,times,twocolumn,authoryear]{elsarticle}

\usepackage{amssymb}
\usepackage{lipsum}
\usepackage{hyperref}
\usepackage{url}
\usepackage{subcaption}
\usepackage{longtable}

\journal{New Astronomy}

\begin{document}

\begin{frontmatter}



\title{A comparative study of occurrence rates and nature of Ultraluminous X-ray sources in spiral and elliptical galaxies}


\author[first]{C. M. Sariga}
\author[first]{P. Shalima}
\ead{shalima.p@manipal.edu}
\author[first]{D. Bhattacharya}
\author[second]{Vivek K. Agrawal}

\affiliation[first]{organization={Manipal Centre for Natural Sciences},
            addressline={Manipal Academy of Higher Education}, 
            city={Manipal},
            postcode={576104}, 
            state={Karnataka},
            country={India}}

\affiliation[second]{organization={Space Astronomy Group},
            addressline={U R Rao Satellite Center, ISITE Campus, Outer Ring Road, Karthik Nagar}, 
            city={Bengaluru},
            postcode={560037}, 
            state={Karnataka},
            country={India}}

\begin{abstract}
Ultraluminous X-ray sources (ULXs) are mostly extragalactic non-nuclear point sources having X-ray luminosity exceeding the Eddington luminosity of 10 $M_\odot$ black hole i.e., $L_X \geq $ 10$^{39}$ erg ~s$^{-1}$. They are observed in all types of galaxies; spirals, ellipticals and dwarf irregulars. But the rate of occurrence of ULXs per galaxy varies, some might host a single ULX, whereas some host a large number. In this work we attempt to identify possible differences in ULX properties between two extreme categories in spirals and ellipticals, i.e. ULXs occurring at a rate of one per galaxy ($N=1$) and those occurring at larger rate. We adopt an effective scheme to generate flux limited, credible samples corresponding to the two groups in spirals and ellipticals. From this study, we infer the presence of a separate population of ULXs in the $N=1$ spiral group which contains a reasonable fraction of both soft and hard sources, while the remaining categories contain mostly harder sources. We also find six ULXs in $N=1$ ellipticals with globular cluster association. In addition, we identify few luminous candidates likely hosting massive accretors. This study provides crucial hints of a potential link between ULX types and their occurrence rates and host morphology, a finding that warrants validation via targeted observations and detailed spectral analysis of these sources.
\end{abstract}




\end{frontmatter}




\section{Introduction}
\label{introduction}
Ultraluminous X-ray sources (ULXs) are extragalactic off-nuclear point sources, which on assuming isotropic emission have X-ray luminosity, ($L_X$) exceeding the Eddington luminosity, {$L_{Edd}$} of an accreting 10 M$_{\odot}$ black hole i.e., $L_X \geq 10^{39}erg~s^{-1}$  \citep{Swartz_2004}. ULXs were first discovered with Einstein observatory \citep{Fabbiano_1989} and thereafter the population size increased with the observations of X-ray missions like {\it ROSAT}, {\it Chandra}, {\it Swift} and {\it XMM-Newton}. Presently, the widely accepted theories to explain the high luminosity for these astrophysical objects includes;  stellar mass black hole binaries accreting at super Eddington limit (\citealp{King_2001}), or a binary with the compact object as rotating and magnetized neutron stars (NS) \citep{Bachetti_2014} known as pulsating ULXs (PULXs) producing beamed emission or geometrical beaming of the radiation \citep{King_2001} emitted from the central regions of a supercritical accretion disk \citep{Abramowicz_1988}. A small fraction of ULX could also be massive black hole candidates in the range $10^2 - 10^4$ M$_{\odot}$ known as intermediate mass black holes (IMBH) accreting at sub-Eddington limit \citep{Remillard_2006}. Within the class of ULXs there exist sources which have $L_X \geq 10^{41}erg~s^{-1}$, categorized as hyperluminous X-ray sources (HLXs) which are ideal sources to search for IMBHs. ULX studies are also important in understanding the extreme accretion regimes, which are vital in interpreting the merger of compact object that are detected as short gamma-ray bursts and gravitational wave sources (\cite{Kovlakas_2020} and references therein).\\

X-ray spectra of a majority of ULXs are characterized by soft excess and high energy curvature. Studies show that hard X-rays may indeed originate in a Comptonizing corona around the inner regions of the accretion disc(\cite{Middleton_2011a}) and the soft component likely originates in accretion disc around compact object or the photosphere at the base of a massive radiatively driven wind. Most of the sources with this spectra have been modeled by a cool disc component plus a power-law having a curvature above $\sim3$ keV \citep{Gladstone_2009}. It is the high energy curvature and its justification of possible origin that can be used to distinguish between black hole ULXs and PULXs. \cite{Pinto_2017} compares both models,  BH-model (diskbb+compTT) with pulsator model (bbody+highecut*powerlaw) and concludes that differentiation is difficult based on the soft component. On the other hand the hard component is expected to originate from the comptonization in an accretion disk corona in the case of BHs and from  magnetically funneled accretion column in the case of NSs. Even though observing pulsations are primary to assign an accretor as NS there can be unobserved pulsations buried in objects where pulsations are not been detected. This is demonstrated in \cite{Walton_2018} by comparing the ULX spectra with that of known PULXs like M82 X-2, NGC 7793 P13, and NGC 5907 ULX. They explain the lack of detected pulsations as being due to the non-pulsed emission from the accretion flow beyond the magnetosphere being larger in comparison to that associated with the accretion columns emission in PULXs. \\

Studies based on {\it XMM-Newton} spectra of ULXs by \cite{Gladstone_2009}, identified three different spectral types ; broadened disc (BD), hard ultraluminous (HUL) and soft ultraluminous (SUL). In BD type ULX, the broad-disc like spectra appears at the Eddington limit and as the accretion rate becomes super-Eddington, initially a higher energy component arises and further increase in accretion rate produces  hard and soft ultraluminous states \citep{Sutton_2013}.\\

ULX are seen in all types of galaxies, but they seem to favour star forming galaxies and the more luminous ones are seen in spirals and irregulars compared to ellipticals (see Reviews by \citealt{Soria_Feng_2011}, \citealt{Tranin_2024}, \citealt{Kaaret_2017} \& \citealt{King_2023}). \cite{Tranin_2024} found the rate of ULX in galaxies to be positively correlated with stellar mass and star formation rate (SFR). They also show a preference for low metallicity or tidally interacting hosts with triggered star formation and are associated with low mass XRB (LMXB-older population) in ellipticals and high mass XRBs (HMXB-younger population) in spirals \citep{Kovlakas_2020}. 

Studies show the possibility of  formation of black holes of $\sim$ 1,000 $M_\odot$ in the densest inner regions of globular clusters with most or all ejected from the cluster by stellar interactions \citep{Maccarone_2007}. \cite{Irwin_2004} describes that, smaller number of ULXs are located in elliptical galaxies and bright ULXs, with $L_X >2\times10^{39} erg s^{-1}$, are extremely rare in elliptical galaxies with some known to be in globular clusters of ellipticals \citep{Maccarone_2007}. \cite{Haaften_2019} suggests that bright ULXs in old populations by means of having a massive BH accretor, may be different from less bright ULXs in the same environments and  their possible origin and properties will be different from the comparable ones in young environments. \cite{Dage_2019} shows a bimodal distribution in  GC ULXs; with disc temperatures (kT) well below 0.5 keV, or variable temperatures ranging above 0.5 keV up to 2 keV. They also noted that the two lowest kT sources shows optical emission lines.

Even though there have been several earlier studies on ULX and their host galaxy dependence, an in depth analysis of a possible connection between ULX-type and their occurence rate in galaxies is limited. 
\cite{Kovlakas_2020} quantified the number of ULXs per galaxy as a function of morphology, SFR, stellar mass ($M_{*}$) and metallicity of their host galaxies. Their results show that $30\%$ of galaxies in all morphological types host sources with luminosities above 
$10^{39} erg\ s^{-1}$ with slightly higher ULXs ($\sim40 \%$) in elliptical galaxies and Sb-Scd type spiral galaxies. However, sources with $L_X >5\times\ 10^{39} erg\ s^{-1}$  are typically hosted by late type galaxies. The number of ULX per galaxy are comparable for early spirals (S0/a-Sb), while in late spirals (Sbc-Sd) they are $\sim$1.5 times higher than for early spirals. This is slightly lower than that of the total population in elliptical galaxies (E). Also, early spirals (S0/a-Sb) exhibit the lowest numbers of ULXs per SFR.\\

It has been observed that ULX count per galaxy varies but there are few detailed studies on the possible reason for some galaxies to host just a single ULX, while others host a larger number and their dependence on host galaxy. More importantly, is there any difference in the nature and origin among ULXs in these groups or their location in the host galaxy or do all have similar properties; these  are some of the interesting questions that initiated this study. \\
In this work, we aim to  explore the dependence of ULX rate on host properties such as, morphology and location in the host, as well as source properties such as $L_X$ and hardness ratio for a sample of ULX with two extreme rates, $N=1$ \& $N\geq7$ in spirals and ellipticals. In order to choose a flux limited and reliable sample, we have employed an alternative approach, the details of which are explained in section-\ref{sec:reliable_method}. Section-\ref{sec:host} describes the host dependence of these ULX groups and section-\ref{sec:ulx} compares the ULX properties to identify possible clustering of ULXs in the $L_{hard}$ vs. $L_{soft}$ plane. We present our conclusions in section-\ref{sect:conclusion}.

\section{ULX sample selection}
\label{sec:sample}
\begin{figure}[hbt!]
\centering
\includegraphics[width=0.3\textwidth]{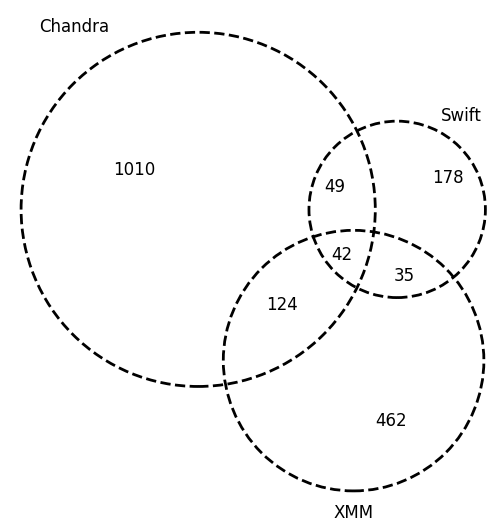}
\caption{Venn diagram representing the number of ULXs with data from {\it Chandra}, {\it XMM-Newton} and {\it Swift} as per the \citet{Tranin_catalog} catalogue.}
\label{fig:ulx_n.png}
\end{figure}

ULX catalogues that contain parameters like hardness ratio, X-ray luminosity, variability index, host properties etc. for different samples are available (\citealt{Swartz_2004}, \citealt{Earnshaw_2019}, \citealt{Bernadich_2022}, \citealt{Tranin_catalog}). For the current study, we utilize the latest catalogue of \citealt{Tranin_catalog} consisting of ULX observed by {\it Chandra}, {\it XMM-Newton} and {\it Swift} missions (see fig. \ref{fig:ulx_n.png}). As the probabilty of some of the X-ray sources to be foreground or background contaminants are high, possible contaminants have been removed from the catalogue, so as to obtain a clean ULX sample (see \citep{Tranin_catalog}). To include more ULXs and study their spatial distribution to large radii, the major axis of the galaxies in this catalogue was considered to be Holmberg diameter $D_{Holm}=1.26\times D_{25}$, ie. ULXs within galaxy diameter of 1.26 times the 25 mag isophote.

For this study, we have only selected the {\it Chandra} and {\it XMM-Newton} observations considering the better resolution and larger sample size respectively. But as the observational energy band for these instruments
are different, 0.5-7.0 keV for {\it Chandra} and 0.2-12.0 keV for {\it XMM-Newton}, for further analysis {\it XMM-Newton} band fluxes were converted to {\it Chandra} band assuming a photon-index of 1.7. As our objective is to study the dependence of ULXs on their host galaxy, ULXs with unknown host or undefined hubble type (T-value) are removed from the sample. Further, all sources with undefined flux values in their detection band are filtered out. The remaining sample is divided into spiral and elliptical categories based on the hubble type T-value, $T\geq0$ and $T<0$ (see Table \ref{tab:ulx_host_n}). For these sets, further classification was done based on the occurence of ULXs per galaxy as shown in Tables \ref{tab:ulx_grps_n_chandra} and  \ref{tab:ulx_grps_n_xmm}.

\begin{table*}[hbt!]
\begin{center}
\caption[]{Total number of ULXs and its hosts.}
\label{tab:ulx_host_n}
 \begin{tabular}{clclclclcl}
  \hline\noalign{\smallskip}& 
  &Spirals (T\(\geq\)0)      & Ellipticals (T $<$ 0) & Total       \\
  \hline\noalign{\smallskip}
{\it Chandra} sample & Number of ULXs& 616 & 312     & 928  \\ 
&Number of hosts& 351     &   211    & 562                \\   \noalign{\smallskip}\hline
{\it XMM-Newton} sample &Number of ULXs& 421 & 166     & 587  \\ 
&Number of hosts& 315     &   133    & 448                \\   \noalign{\smallskip}\hline
\end{tabular}
\end{center}
\end{table*}

\begin{table}[hbt!]
\begin{center}
\caption[]{Total number of ULXs in each group for {\it Chandra} sample}
\label{tab:ulx_grps_n_chandra}
\begin{tabular}{ccccc}
 \hline\noalign{\smallskip}
   {Groups} &
    \multicolumn{2}{c}{Number in Spirals} &
    \multicolumn{2}{c}{Number in Ellipticals} \\
& ULXs & Hosts & ULXs & Hosts  \\
  \hline\noalign{\smallskip}
$N=1$&235 &235 & 167 &167\\  
$2 \leq N \leq 3$& 204&86&83&35 \\    
$4 \leq N \leq 9$ & 150 &28 & 42&8 \\ 
$N \geq 10 $& 27 &2& 20&1 \\
  \noalign{\smallskip}\hline
\end{tabular}
\end{center}
\end{table}

\begin{table}[hbt!]
\begin{center}
\caption[]{Total number of ULXs in each group for {\it XMM-Newton} sample}
\label{tab:ulx_grps_n_xmm}
\begin{tabular}{ccccc}
 \hline\noalign{\smallskip}
   {Groups} &
      \multicolumn{2}{c}{Number in Spirals} &
      \multicolumn{2}{c}{Number in Ellipticals} \\
     & ULXs & Hosts & ULXs & Hosts       \\
  \hline\noalign{\smallskip}
$N=1$&249 &249 & 111 &111\\  
$2 \leq N \leq 3$& 129&57&38&18 \\    
$4 \leq N \leq 9$ & 43 &9 & 17&4 \\ 
  \noalign{\smallskip}\hline
\end{tabular}
\end{center}
\end{table}

But in this sample, there exists a possibility for larger number of ULXs being detected in a particular galaxy since it is nearby or its angular size is large. To remove such possible observational biases caused by distance and extent of galaxy, and to ensure that most of the ULXs are observed, we have created a subsample based on characteristics of a galaxy hosting large number of ULXs (in our sample Cartwheel galaxy with $N\geq$10 ULXs). Therefore, our sample is constrained to have a semi-major axis of at least 37.8$''$ and distance of at most 126.5 Mpc. ULXs were regrouped after this sample filtration and our target groups for study were chosen as ULX with two extreme rates, $N=1$ \& $N\geq7$ in spirals and ellipticals (sample details in Table \ref{tab:ulx_grps_bias_n}). $N\geq7$ was particularly selected instead of $N\geq10$ because in elliptical category, $N\geq10$ is hosted by Cartwheel galaxy which has a ring type morphology hence not considered for further analysis and the second largest sample size was 7. Hence this limit was considered for the generating the sample of ULXs occurring at larger rate. 

\begin{table*}[hbt!]
\begin{center}
\caption[]{Total number of ULXs after observational bias correction }
\label{tab:ulx_grps_bias_n}
 \begin{tabular}{clcccc}
  \hline\noalign{\smallskip}
  &{Groups} &
      \multicolumn{2}{c}{Number in Spirals} &
      \multicolumn{2}{c}{Number in Ellipticals} \\
    & & ULXs & Hosts & ULXs & Hosts       \\
  \hline\noalign{\smallskip}
{\it Chandra} sample& $N=1$ & 115 &115&53& 53\\ 
&$N \geq 7 $& 49&4 & 14&2 \\
{\it XMM-Newton} sample & $N =1 $& 112 &112& 46&46\\
&$N\geq7$&8&1&-&-\\
  \noalign{\smallskip}\hline
\end{tabular}
\end{center}
\end{table*}

For the detailed analysis, we have considered the flux values for {\it Chandra} sample from the \citet{Chandra_catalog} catalogue, which provides the average value of aperture-corrected net energy flux inferred from the source region from one or more observations. Similarly, the flux values for {\it XMM-Newton} sample are taken from \cite{XMM_catalog}. Flux uncertainties are obtained from the respective catalogues and the uncertainties in luminosity are calculated using error propagation formula \citep{Bevington_2003}.
 Sources with counts above 5$\sigma$ were selected from each group. As there is a possibility of multi-ULX hosts to be misclassified into the $N=1$ category due to limited exposure, it is important to ensure that no fainter ULXs have been missed in the sample. Therefore, in order to generate a credible $N=1$ sample, we have adopted an effective framework as illustrated in the following section-\ref{sec:reliable_method} . 

\subsection{Creating a credible $N=1$ sample }
\label{sec:reliable_method}
To ensure the reliability of the generated $N=1$ group, we determine the limiting flux for 5$\sigma$ cts. The methodology is outlined below;
\begin{enumerate}
    \item Source count rate ($CPS_{src}$) for energy flux in 0.5-7.0 keV band is calculated using \href{https://heasarc.gsfc.nasa.gov/cgi-bin/Tools/w3pimms/w3pimms.pl}{WebPIMMS}\footnote{\url{https://heasarc.gsfc.nasa.gov/cgi-bin/Tools/w3pimms/w3pimms.pl}} for the entire sample assuming a power-law model.
    (Caveat : CPS estimation with power-law model of index 1.7 and considering $nH$=0)
    \item Source counts ($CTS_{src}$) were calculated using the total exposure for all observations in detecting a source which is available in the respective catalogues $t_{exp}$, \[CTS_{src}=CPS_{src}\times t_{exp}\]
    \item Sources with counts above 5$\sigma$ are selected
    \item Count rate corresponding to 25 cts is calculated at the observed exposure time
    \item Ratio of count rate corresponding to 25 cts to the source count rate is calculated as, \[\alpha = \frac{CPS_{25}}{CPS_{src}}\]
    \item As energy flux is proportional to count rate for a given instrument we get,
    \[\frac{CPS_{25}}{CPS_{src}}=\frac{F_{lim}}{F_{src}}\]
    \item Then the flux corresponding to 25 cts, which is the limiting flux considererd for the sample is given by,
    \[F_{lim}=\alpha \times F_{src} \]
    \item Limiting luminosity, $L_{Xlim}$ for a source at distance $D$, is given by \[L_{Xlim}=4\pi D^2\times F_{lim}\]
    \item Limiting luminosity, $L_{Xlim}$ $\leq 10^{39}erg~s^{-1}$  are considered to be reliable $N=1$ sources, as no ULXs fainter than the observed one would be excluded
    \item Further sources were also cross-matched with other missions to filter out hosts with multiple ULXs
\end{enumerate}

With this method, we define the luminosity cut for the host to detect faintest ULXs at the given exposure time. If the $L_{Xlim}$ exceeds threshold limit of ULX $\sim 10^{39}erg~s^{-1}$, there exists a chance for less luminous ULXs to be hosted by the galaxy other than the observed one. Whereas, if $L_{Xlim} \leq 10^{39}erg~s^{-1}$, no other faint ULXs have been missed out at this given exposure time, making it a truly credible sample of $N=1$ ULXs. Initial sample size, size after 5$\sigma$ filteration and reliable sample size are shown in Table \ref{tab:ulx_fiteration_n}. The final datasets are given in \ref{sec:elliptical_dataset} (elliptical category ) and \ref{spiral_dataset} (spiral category).

For sources observed by both {\it Chandra} and {\it XMM-Newton}, only {\it Chandra} observations are retained and hence the final sample size for our target groups is summarized in Table \ref{tab:ulx_final_n}.

\begin{table*}[hbt!]
\begin{center}
\caption[]{Complete sample details of ULX groups }
\label{tab:ulx_fiteration_n}
 \begin{tabular}{clcc}
  \hline\noalign{\smallskip}
&Groups      & Number of ULXs in Spirals &Number of ULXs in Ellipticals    \\ 
  \hline\noalign{\smallskip}
Chandra sample& $N=1$ & 115 & 53\\ 
&$\geq5\sigma$&37 &11\\
&Reliable & 22 & 9\\
&$N \geq 7 $&49 &14\\
&$\geq5\sigma$&19 &6\\
XMM sample & $N=1$& 112 & 46\\
&$\geq5\sigma$&79&36\\
&Reliable & 53 & 25\\
&$N \geq 7 $& 8&-\\
&$\geq5\sigma$& 8&-\\
  \noalign{\smallskip}\hline
\end{tabular}
\end{center}
\end{table*}

\begin{table}[hbt!]
\begin{center}
\caption[]{Final sample of ULX groups }
\label{tab:ulx_final_n}
 \begin{tabular}{clccc}
  \hline\noalign{\smallskip}
  {Groups} &
      \multicolumn{2}{c}{Number in Spirals} &
      \multicolumn{2}{c}{Number in Ellipticals} \\
     & ULXs & Hosts & ULXs & Hosts       \\
  \hline\noalign{\smallskip}
 $N=1$ & 68 &68&30& 30\\ 
$N \geq 7 $& 20&4 & 6&1 \\

  \noalign{\smallskip}\hline
\end{tabular}
\end{center}
\end{table}

 It should be also noted that none of the ULXs crossed the $L_X$ limit of HLX sources, $L_X > 10^{41} erg~ s^{-1}$. This sample can be a heterogeneous population, their accretors can be stellar-mass black holes, neutron stars or IMBHs. Considering $L_X$ distribution and hardness ratios in the following sections we try to identify different types of ULXs occuring at two extreme rates, $N=1$ \& $N\geq7$ in spirals and ellipticals. 

\section{Results and Discussion}

\subsection{Dependence on Host properties}
\label{sec:host}
\subsubsection{Morphology}
\begin{figure}[hbt!]
\centering
\includegraphics[width=0.4\textwidth]{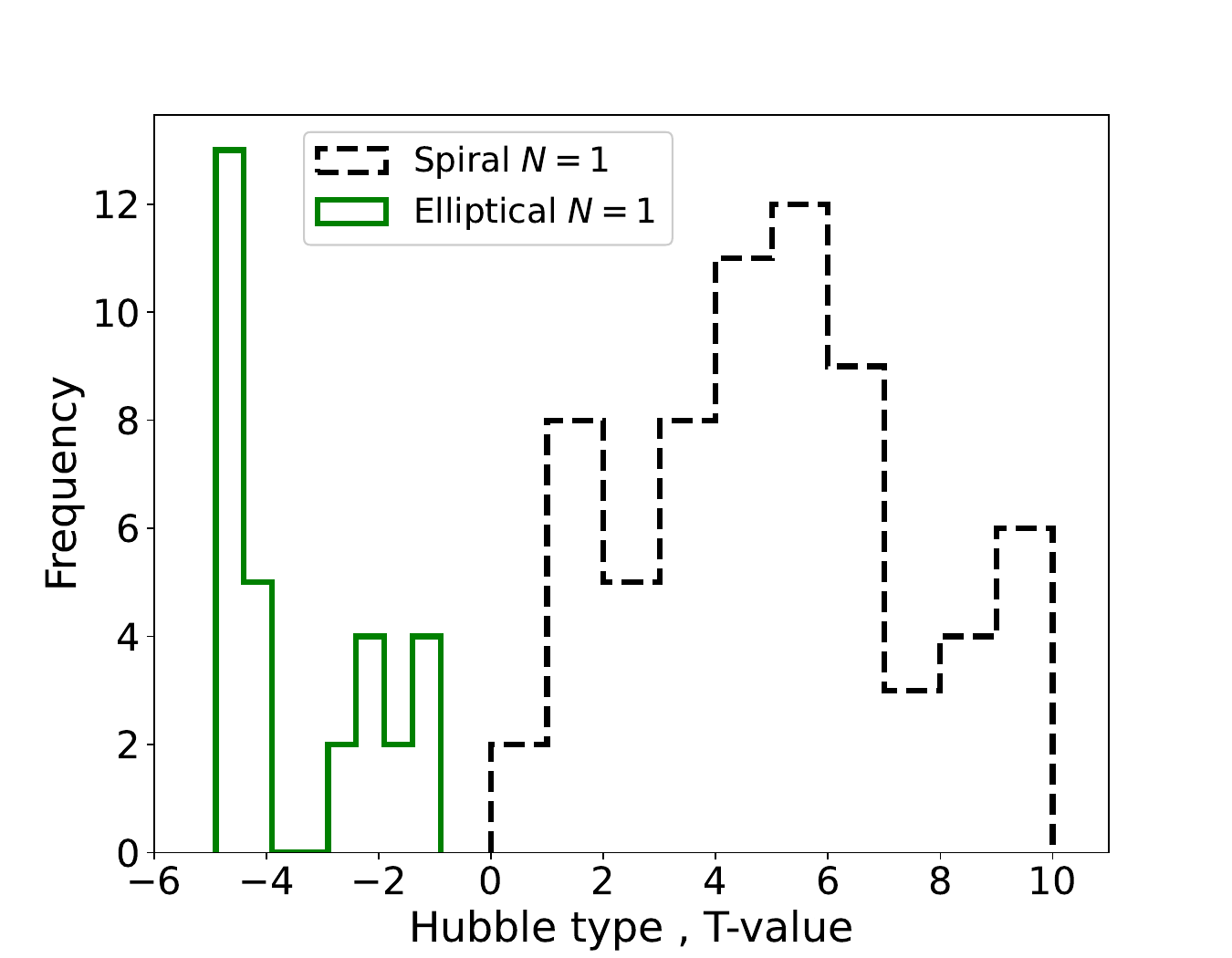}
\caption{Hubble type, T-value distribution of ULXs in spiral and elliptical categories of $N=1$ group}
\label{fig:t_dist}
\end{figure}

The ULX frequency as a function of the Hubble-type, T-values (\cite{hubble_binney_1998}, \cite{deVaucouleurs_1974}) for $N=1$ group is shown in Figure \ref{fig:t_dist}. In De Vaucouleurs galaxy classification scheme, the Hubble T-values range from –6 to +10 with the negative values corresponding to early types and positive values indicating star forming galaxies. The T-value distribution indicates that ULXs are hosted by all types of spirals covering a broader range of T-values. Whereas, ellipticals hosts are primarily peaked around a range of $-5 \leq$ T $\leq -4$ which correspond to normal and late ellipticals, where the late types are star forming elliptical galaxies. The $N\ge7$ category was not considered here due to poor statistics.

\subsection{Identification of ULX types}
\label{sec:ulx}
\subsubsection{X-ray luminosity ($L_X$) distribution}

\begin{figure}[hbt!]
\centering
\begin{subfigure}[b]{0.5\textwidth}
    \centering
    \includegraphics[width=\textwidth]{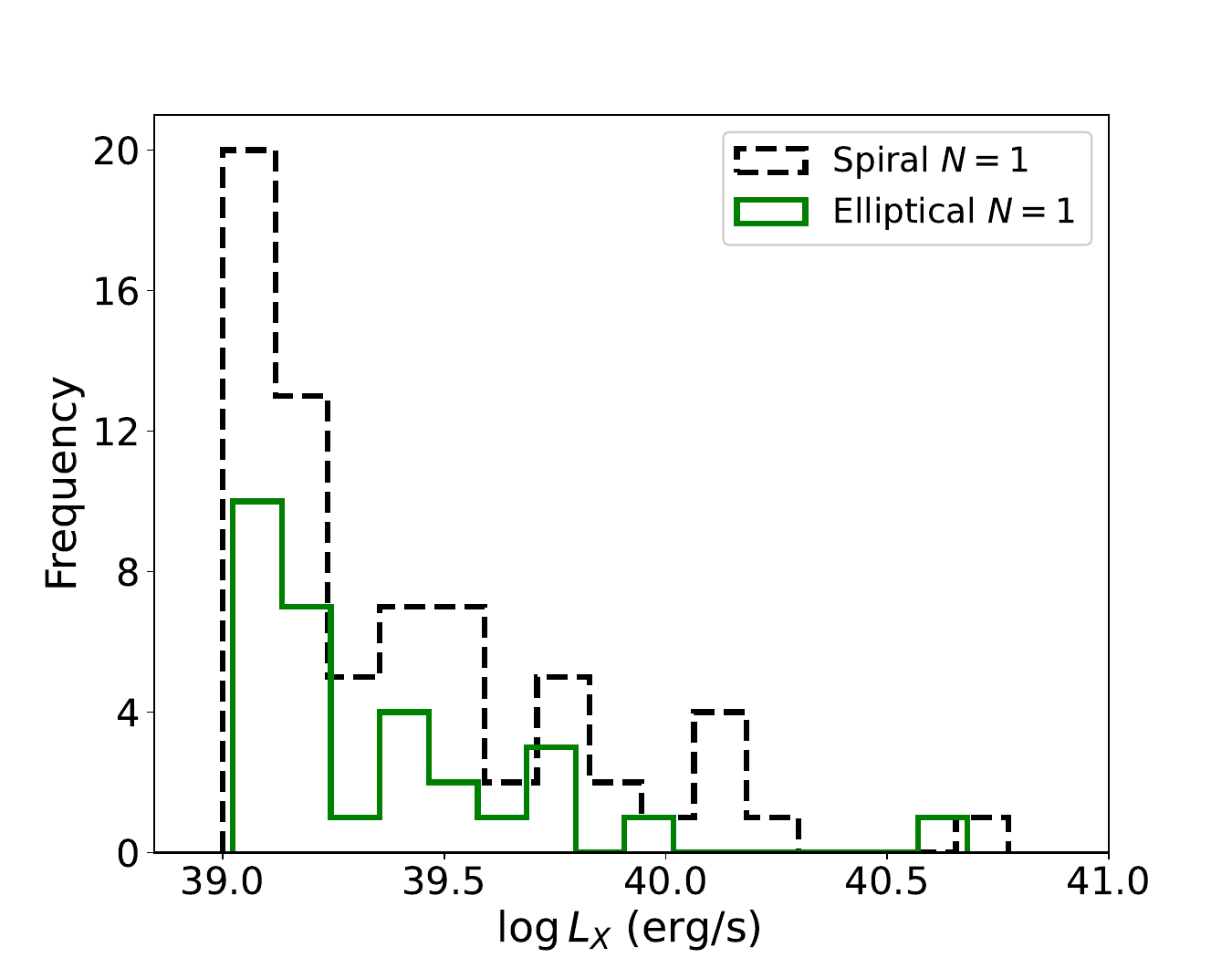}
    \caption{$N=1$ groups}
    \label{fig:lx_1}
\end{subfigure}
\hfill
\begin{subfigure}[b]{0.5\textwidth}
    \centering
    \includegraphics[width=\textwidth]{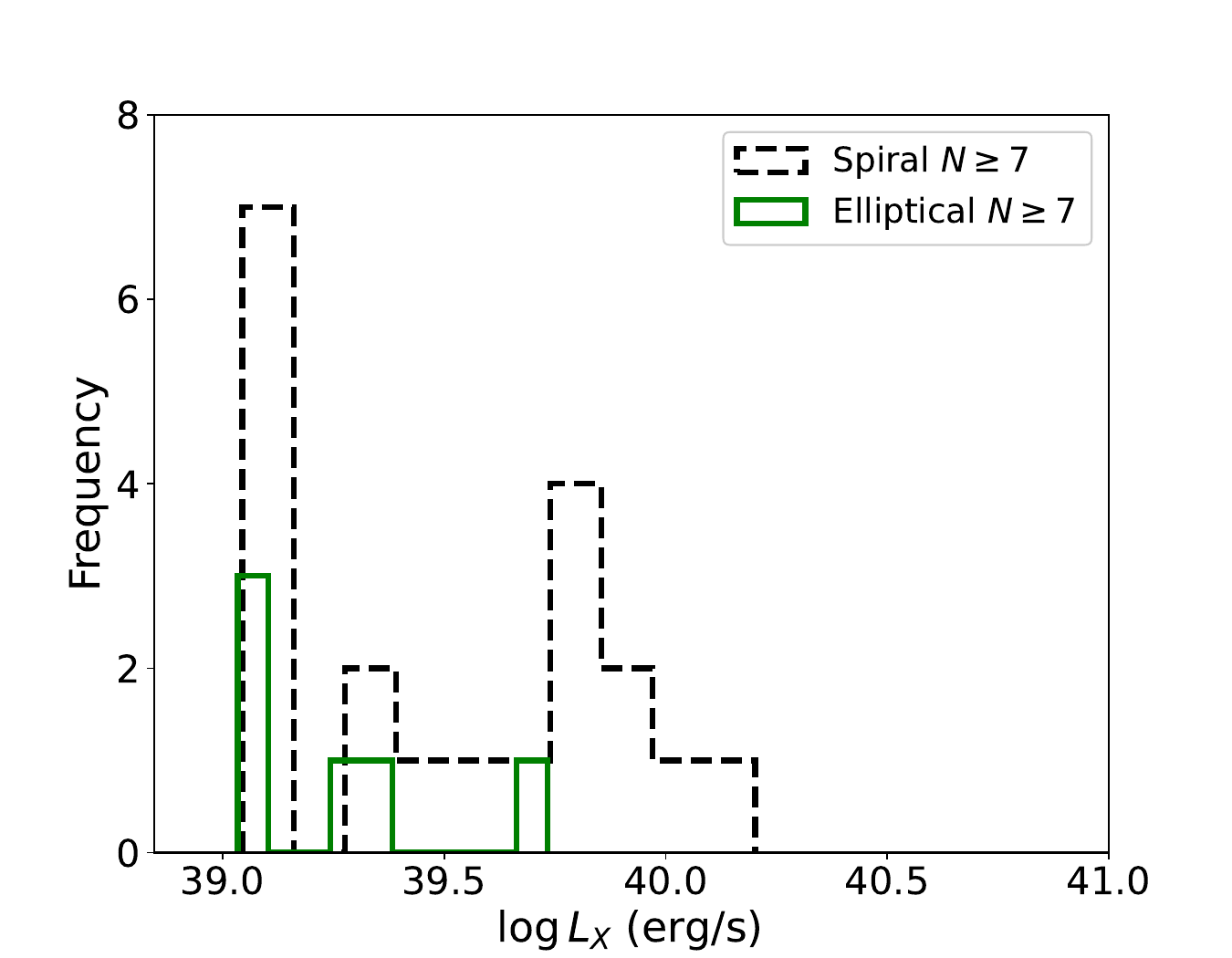}
    \caption{$N\geq7$ groups}
    \label{fig:lx_7}
\end{subfigure}
\caption{$L_X$ distribution of ULXs in spiral and elliptical categories for 
(a) $N=1$ groups and (b) $N\geq7$ groups.}
\label{fig:lx_dist}
\end{figure}

The $L_X$ distribution of elliptical and spiral hosted ULXs in each group is shown in Figure \ref{fig:lx_dist}. It shows that with increase in luminosity, the number of ULXs decreases for both categories as expected. We also infer that, spiral and elliptical hosted sources in $N=1$ category span a similar range of luminosities. However, $N\geq7$ spiral ULXs have a bimodal distribution and have larger fraction of more luminous sources compared to $N\geq7$ ellipticals.
\\
Previous studies show ULXs in ellipticals to be associated with old stellar populations including globular clusters (\cite{Maccarone_2007}, \cite{Kovlakas_2020}, \cite{Dage_2024}) and may be long-lasting transient outbursts in low-mass X-ray binaries (LMXBs) \citep{King_2002}. 
According to \citet{Irwin_2004}, ULXs in early type galaxies are less luminous ($L_X<2\times10^{39}erg~s^{-1}$) than those in late types which can be explained by accretion onto 10 - 20 $M_{\odot}$ black holes and ULXs above this limit are rare in ellipticals. Their study also indicates that ULXs in ellipticals are associated with LMXBs as the elliptical galaxies show flatter trend between ULX frequency and host stellar mass and ULXs in spirals are mostly associated with high-mass X-ray binaries (HMXBs). They also add that LMXBs approach ULX luminosity during X-ray outbursts when  in very high state (\cite{Remillard_2006}). 

In our filtered sample of $N=1$ elliptical category, there exists five ULXs with $L_X \geq 5\times 10^{39}~erg~s^{-1}$ and one above $L_X \geq 10^{40}~erg~s^{-1}$, which may be massive accretors if they are softer sources (see Figure \ref{fig:lx_1}). But all sources in $N\geq7$ elliptical category are less luminous except one with $L_X \sim 5\times 10^{39} ~erg~s^{-1}$. 
Apart from the few brightest sources, majority of our ULXs may correspond to super-critically accreting stellar-mass black holes of few solar mass or neutron stars with low-mass companions.  
\subsubsection{Hardness ratio}
\begin{figure*}[hbt!]
\centering
\begin{subfigure}[b]{0.4\textwidth}
    \centering
    \includegraphics[width=\textwidth]{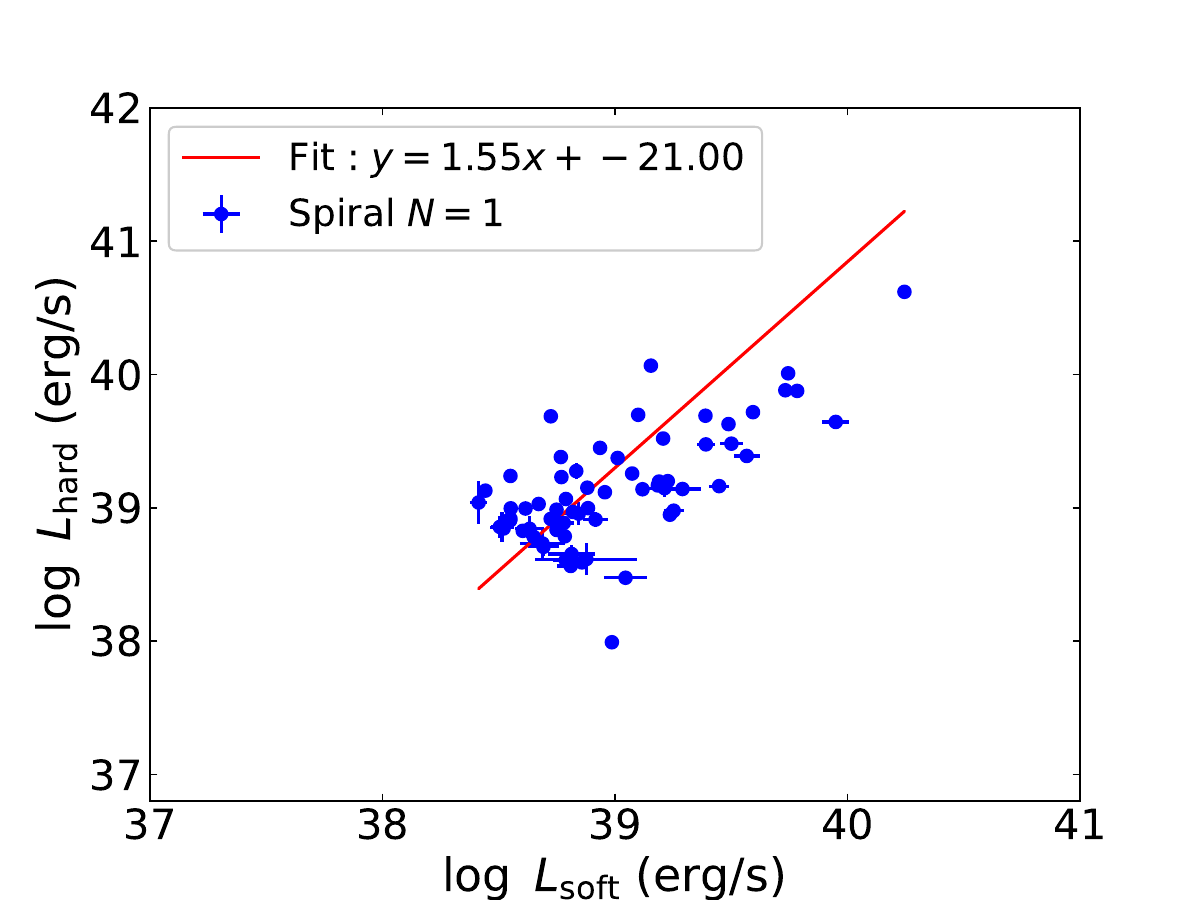}
    \caption{Spirals, $N=1$}
    \label{fig:soft_hard_1a}
\end{subfigure}
\hspace{0.01\textwidth}
\begin{subfigure}[b]{0.4\textwidth}
    \centering
    \includegraphics[width=\textwidth]{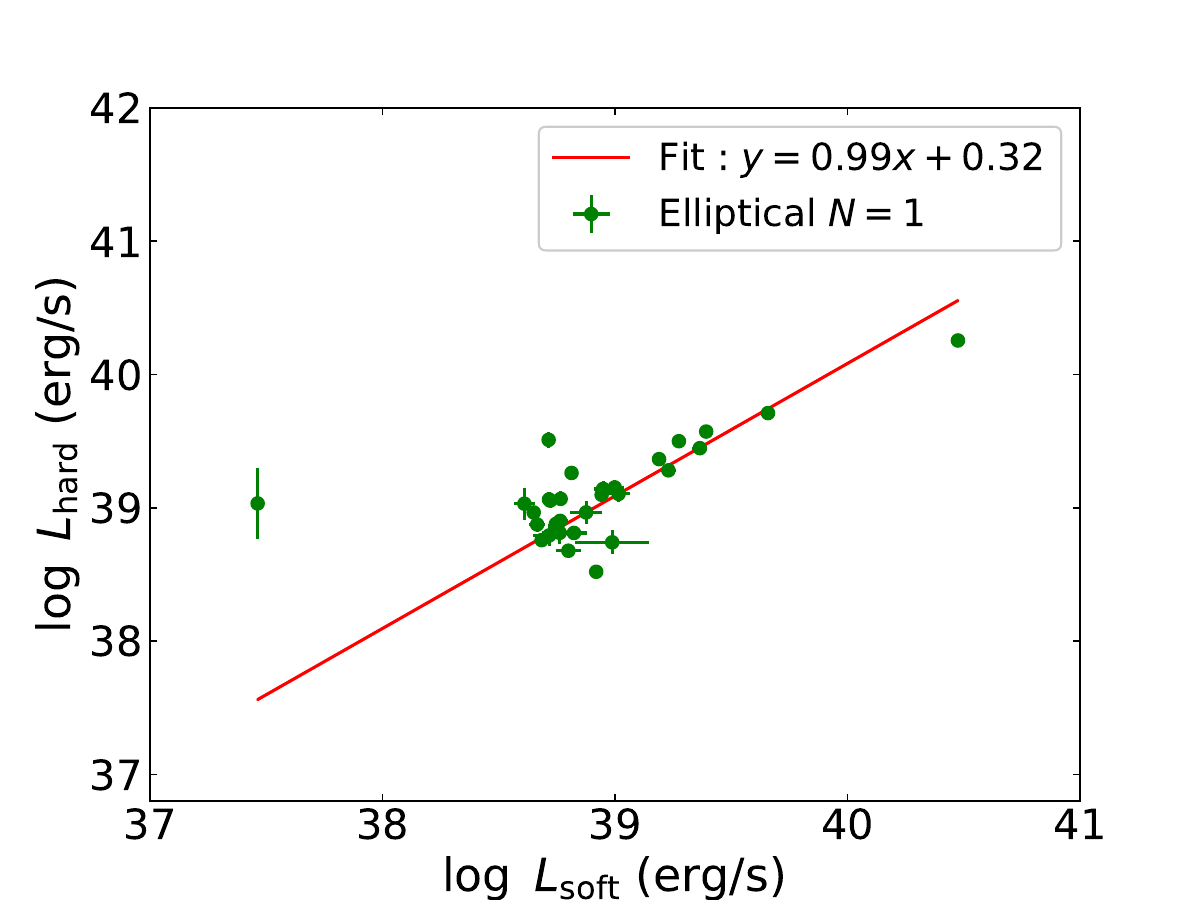}
    \caption{Ellipticals, $N=1$}
    \label{fig:soft_hard_1b}
\end{subfigure}
\vspace{5pt}
\begin{subfigure}[b]{0.4\textwidth}
    \centering
    \includegraphics[width=\textwidth]{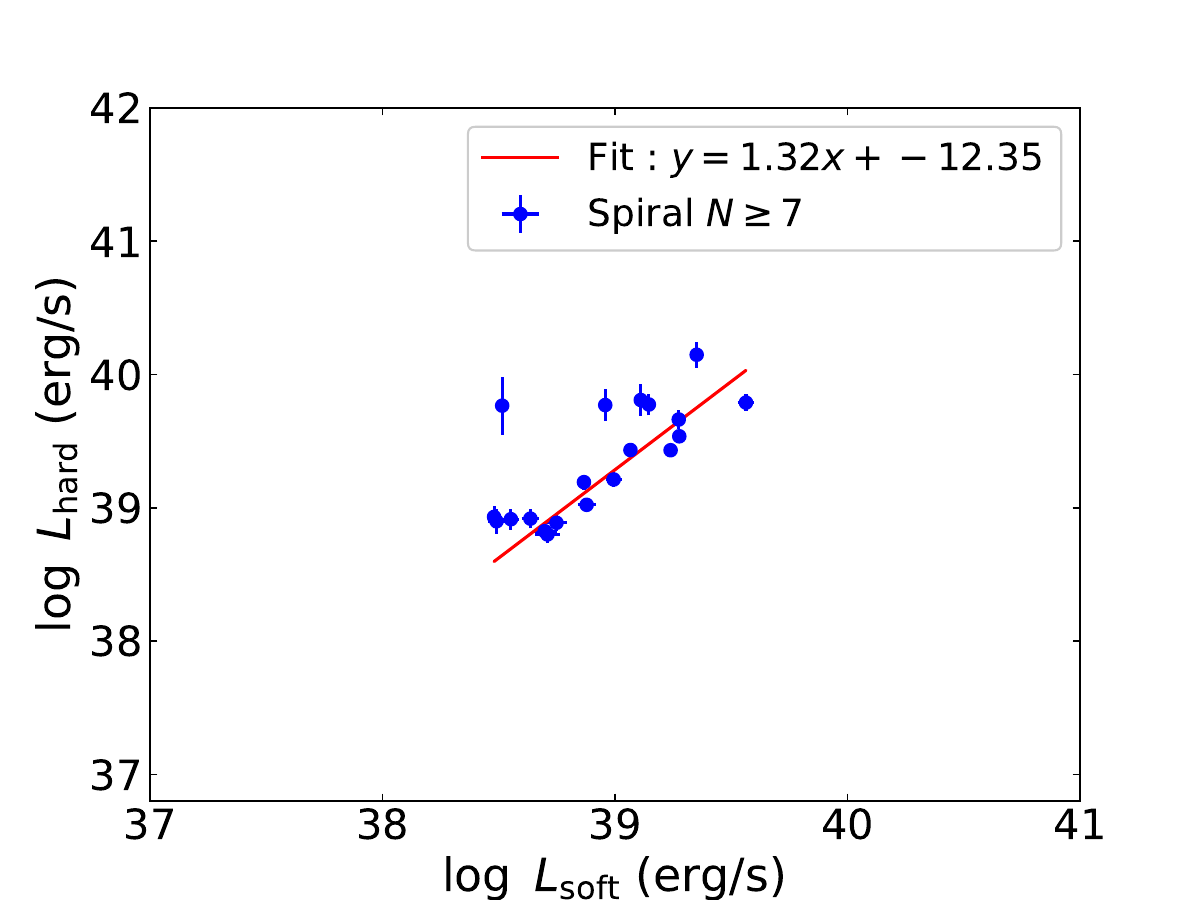}
    \caption{Spirals, $N\geq7$}
    \label{fig:soft_hard_7a}
\end{subfigure}
\hspace{0.01\textwidth}
\begin{subfigure}[b]{0.4\textwidth}
    \centering
\includegraphics[width=\textwidth]{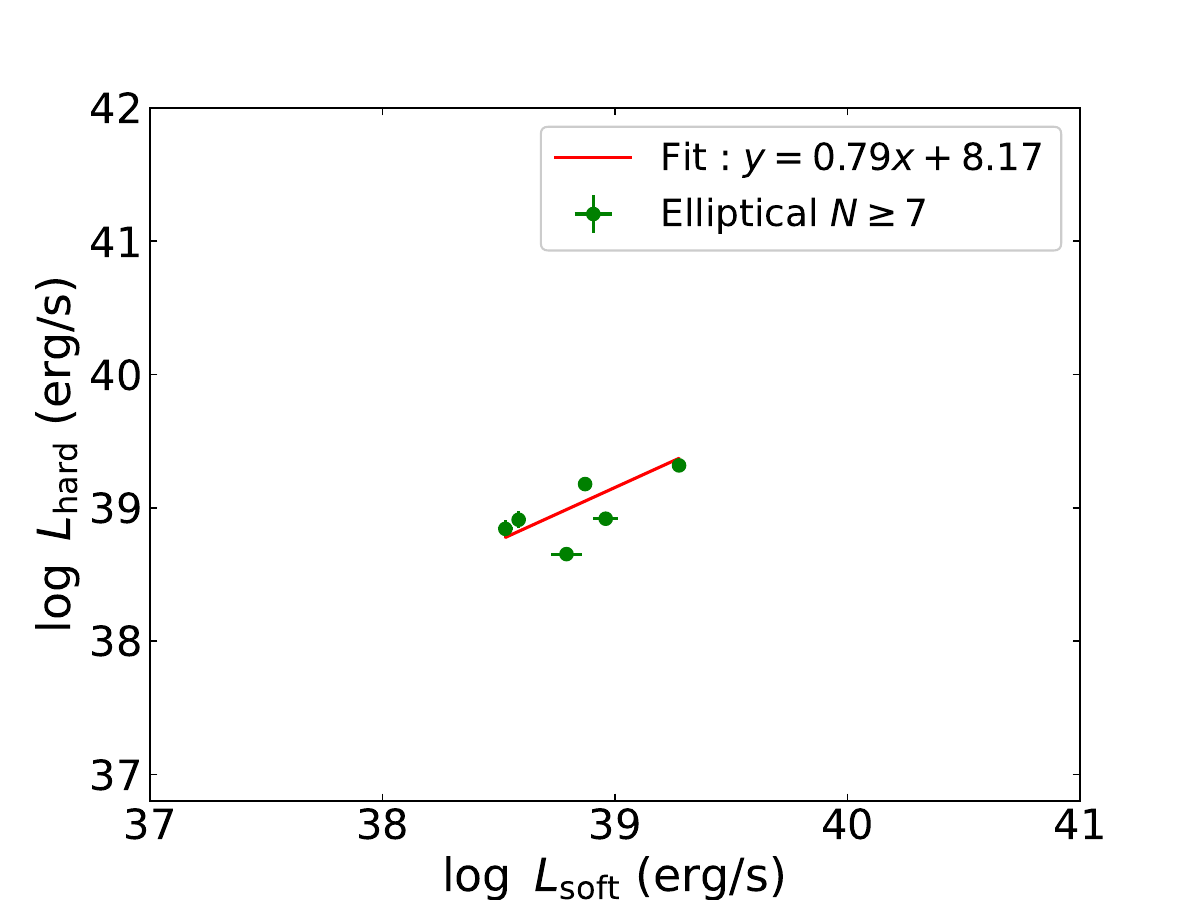}
    \caption{Ellipticals, $N\geq7$}
    \label{fig:soft_hard_7b}
\end{subfigure}
\caption{Hard vs soft X-ray luminosity plots for ULXs in: (a) $N=1$ spirals, (b) $N=1$ ellipticals, (c) $N\geq7$ spirals and (d)  $N\geq7$ ellipticals.}
\label{fig:lhard_lsoft_all}
\end{figure*}

\begin{figure*}[hbt!]
\centering
\begin{subfigure}[b]{0.4\textwidth}
    \centering
    \includegraphics[width=\textwidth]{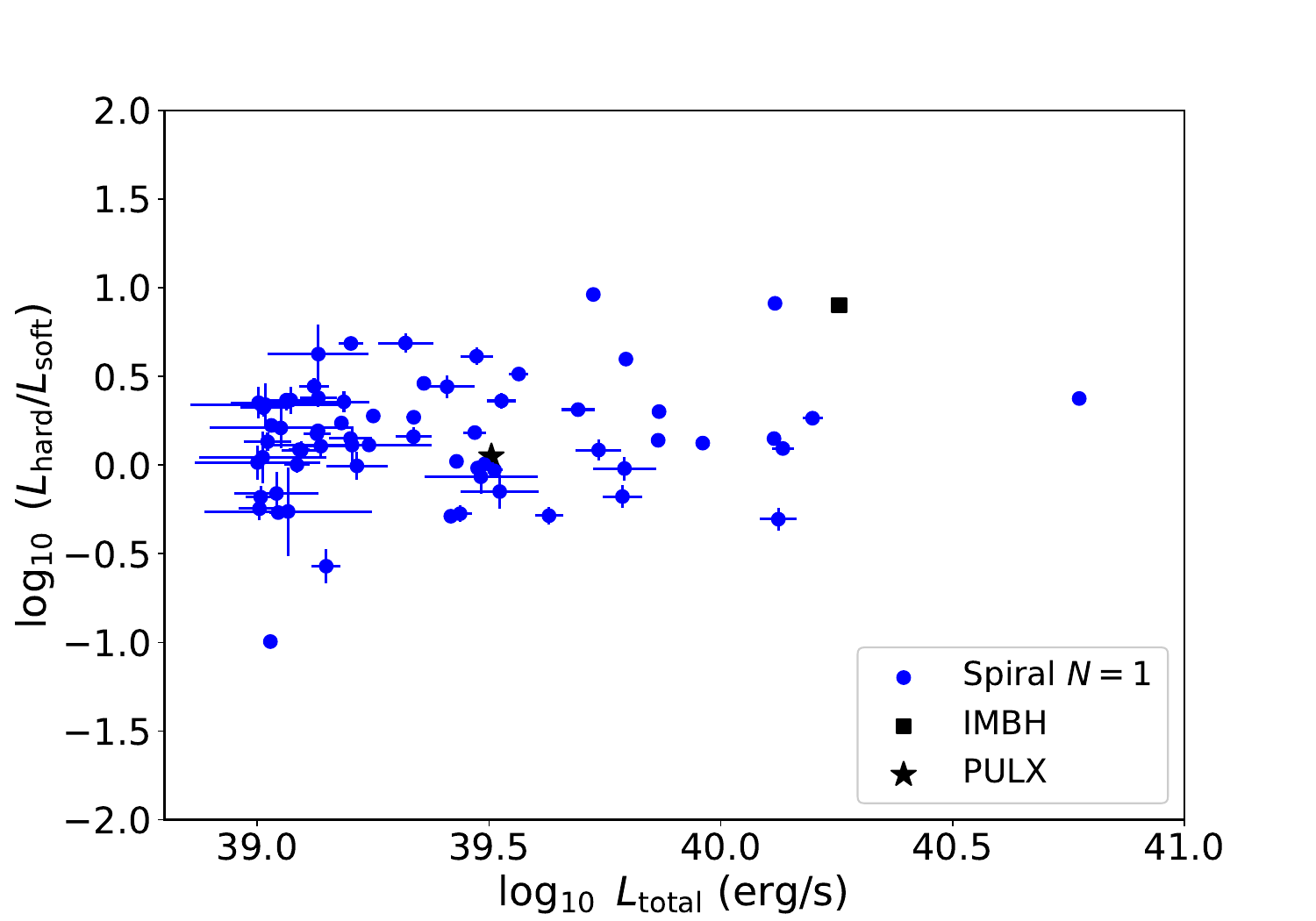}
    \caption{Spirals, N=1}
    \label{fig:ratio_1a}
\end{subfigure}
\hspace{0.01\textwidth}
\begin{subfigure}[b]{0.4\textwidth}
    \centering
    \includegraphics[width=\textwidth]{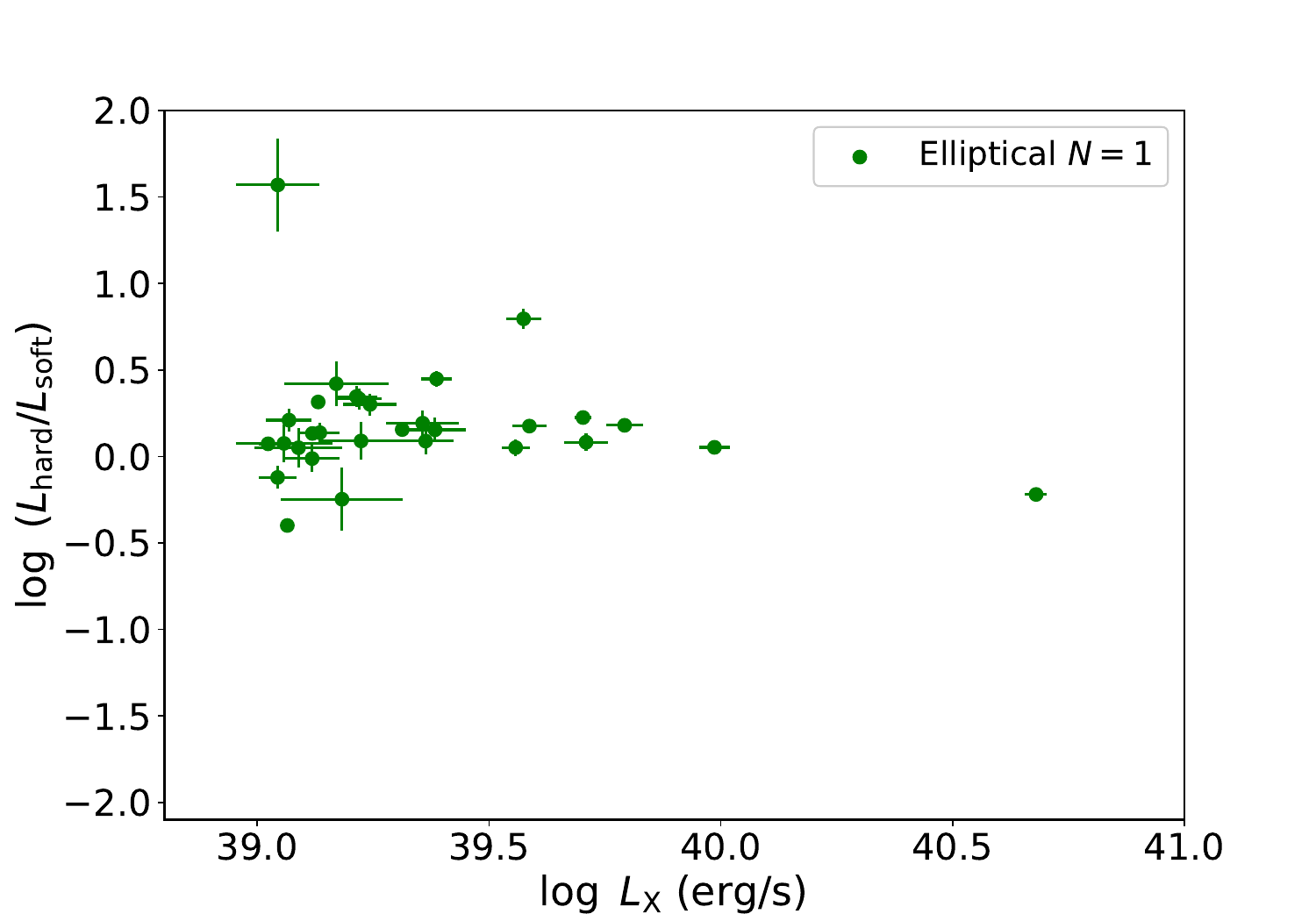}
    \caption{Ellipticals, N=1}
    \label{fig:ratio_1b}
\end{subfigure}
\vspace{5pt}
\begin{subfigure}[b]{0.4\textwidth}
    \centering
    \includegraphics[width=\textwidth]{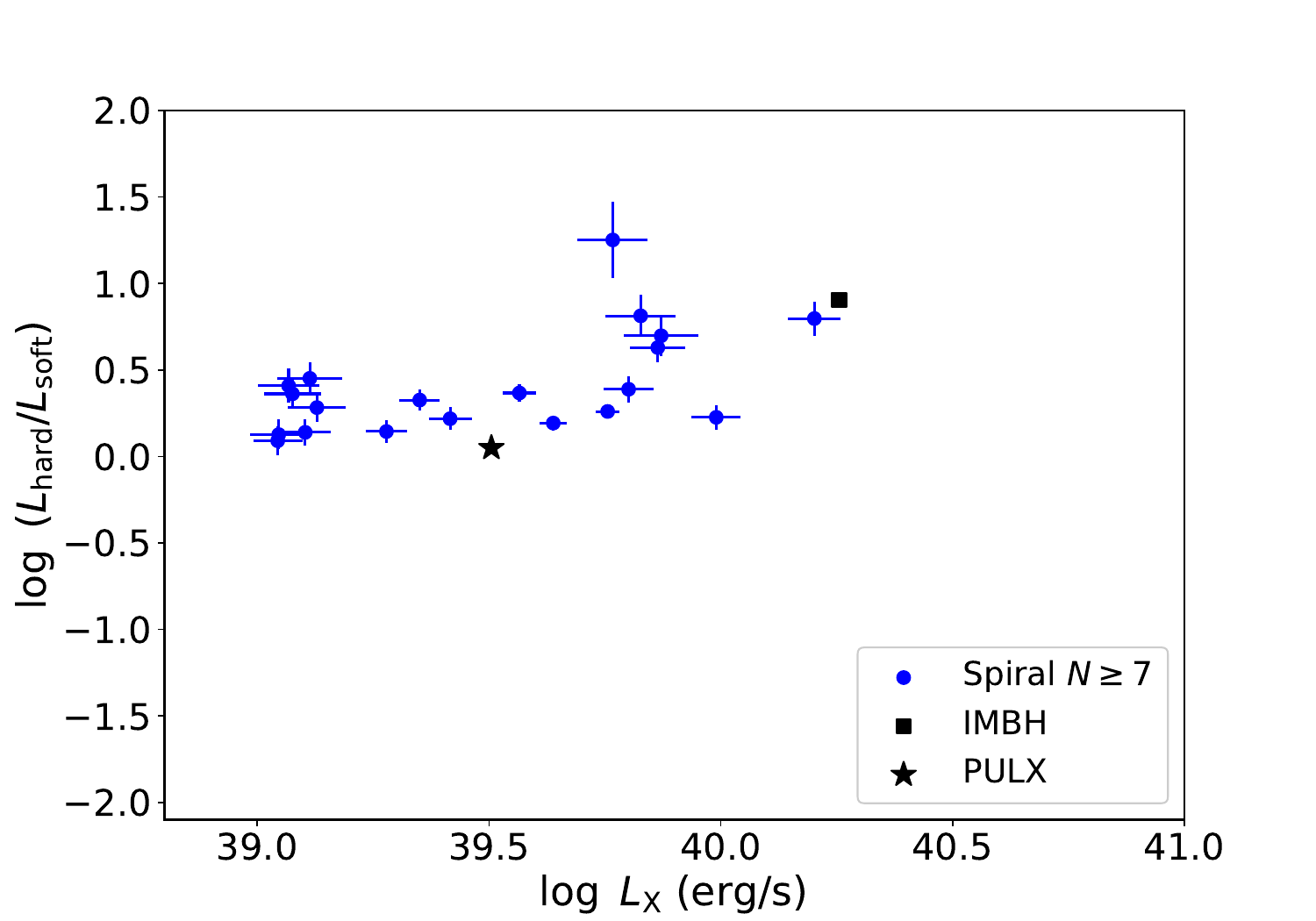}
    \caption{Spirals, $N\geq7$}
    \label{fig:ratio_7a}
\end{subfigure}
\hspace{0.01\textwidth}
\begin{subfigure}[b]{0.4\textwidth}
    \centering
    \includegraphics[width=\textwidth]{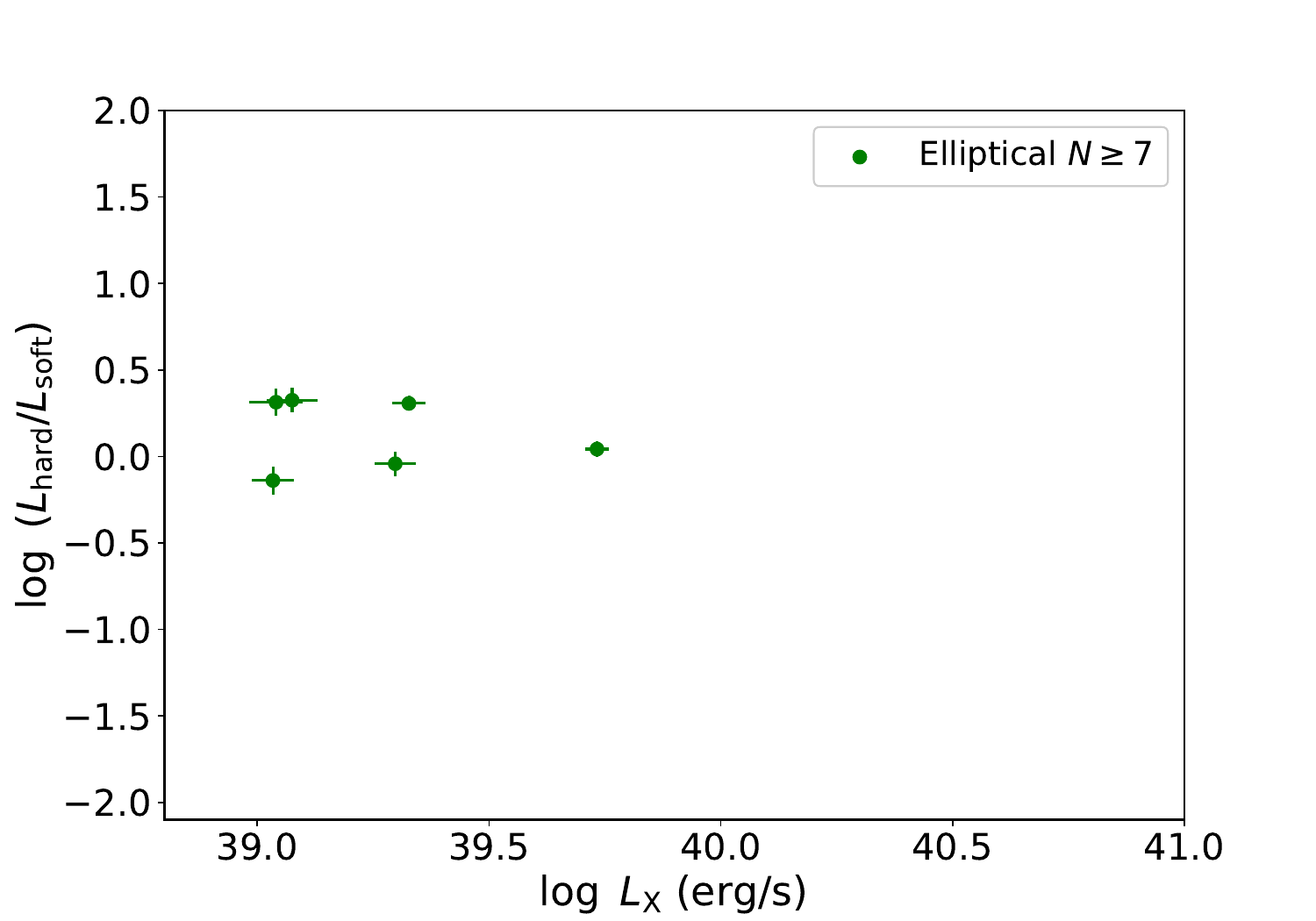}
    \caption{Ellipticals, $N\geq7$}
    \label{fig:ratio_7b}
\end{subfigure}
\caption{Ratio of log $L_{hard}$ to $L_{soft}$ vs. broad-band  X-ray luminosity plot for ULXs in: (a) $N=1$ spirals, (b) $N=1$ ellipticals, (c) $N\geq7$ spirals and (d)  $N\geq7$ ellipticals.}
\label{fig:lx_ratio_all}
\end{figure*}

Luminosity to hardness evolution studies are often used to distinguish between XRBs and ULXs. Hence, we have studied the dependence of hardness ratio (HR) on $L_X$ for our ULX sample (see Figs. \ref{fig:lhard_lsoft_all} \& \ref{fig:lx_ratio_all}). We estimate the hardness ratio as $\frac{L_{hard}}{L_{soft}}$, where hard band is defined in the range 2.0 - 7.0 keV and soft band in the range 0.5 - 2.0 keV. Also, a correlation study of $L_{hard}$ vs. $L_{soft}$ is carried out. 

Figure \ref{fig:lhard_lsoft_all}, shows the plot of  $L_{hard}$ vs. $L_{soft}$ for all categories in our sample. We see that, all samples show a high positive correlation between $L_{hard}$ and $L_{soft}$ with a correlation coefficient greater than 0.75 except for $N\geq7$ spiral category which shows a moderate correlation. A comparison of best fit slope values of $N=1$ spiral ULXs (Figure \ref{fig:soft_hard_1a}) with all other samples shows that, these sources are harder. But for the same category, from the hardness vs. $L_X$ plot (see Figure \ref{fig:ratio_1a}) we find a reasonable fraction of soft ($\sim$ 26.5\%) and hard ($\sim$ 73.5\%) ULXs, with high luminous ($L_X >\ 1 \times 10^{40}\ erg\ s^{-1}$) sources being mostly harder.

We also find that ULXs in $N=1$ spirals are more scattered compared to the other groups and there exist a significant slope variation between sources in galaxy types. Slopes are higher for spirals (1.55, 1.32) and low for ellipticals (0.99, 0.79) indicating a hardness difference. Also, from Figure \ref{fig:lx_ratio_all} we conclude that $N=1$ spiral hosts both soft and hard sources whereas all other groups contain primarily harder sources.

To compare the sample distributions, a Kolmogorov-Smirnov test (K-S test) was conducted using the hardness ratio. Results for samples showing significantly different distributions at the 95\% confidence level are shown in Table \ref{tab:ks_test}. We see that there exists no enough evidence suggest that ULXs in $N=1$ spirals and $N=1$ ellipticals are from different populations. However, ULXs in $N=1$ spiral and $N=1$ elliptical category form a different population compared to the $N\geq7$ spiral group. Due to the smaller sample size of $N\geq7$ elliptical sample, the test results are unreliable hence not considered here. 

\begin{table}[hbt!]
\begin{center}
\caption[]{KS test results }
\label{tab:ks_test}
 \begin{tabular}{ccc}
  \hline\noalign{\smallskip}
Samples &Test statistics& p-value  \\ 
  \hline\noalign{\smallskip}
N=1 spiral and elliptical ULXs& 0.17 & 0.51 \\ 
N=1 spiral and $N\geq7$ spiral ULXs& 0.40& 8.41e-3 \\
N=1 elliptical and $N\geq7$ spiral ULXs & 0.46&7.79e-3 \\
  \noalign{\smallskip}\hline
\end{tabular}
\end{center}
\end{table}

To identify the possible ULX type, we have plotted the location of a known IMBH candidate (NGC 2276-3c) and PULX candidate (NGC 7793 ULX-4) in Figures \ref{fig:ratio_1a} \& \ref{fig:ratio_7a} as both are in spirals. NGC 2276-3c has a hard band luminosity,  $L_{hard} (2.0-10.0~keV)$ of $ 1.6\times10^{40}erg~s^{-1}$  and broadband luminosity,  $L_{total} (0.3-10.0~keV)$ of $ 1.8\times10^{40}erg~s^{-1}$ \citep{Mezcua_2015}. NGC 7793 ULX-4 has $L_{hard} (2.0-10.0~ keV)$ of $1.72\times10^{39}erg~s^{-1}$ and  $L_{total} (0.3-10.0~ keV)  $ of $3.26\times10^{40}erg~s^{-1}$  \citep{Quintin_2021}. 
Also, all the known PULXs are among the hardest ULXs which differentiates these types from BH-ULXs. Studies indicate that XRBs show two canonical states; high soft and low hard, a hysterisis cycle \citep{Remillard_2006}. \cite{Sutton_2013} showed that some of the ULXs tend to soften on brightening which indicated the presence of BHs similar to XRBs. However, PULX like NGC 5907 ULX-1 and few BH-ULXs tend to show harder spectra when they brighten \citep{Gurpide_2021}.  

Most of the luminous sources identified here are harder sources which also matches with the IMBH candidate location from the literature, implying data corresponding to its harder state. However, theoretically luminous and softer sources are ideal candidates for massive accretors i.e., inverse relation of accretor mass with temperature indicates that massive objects are softer sources and hence IMBHs are expected to be softer considering thin disc accretion ($T\propto M^{-1/4}$ also see in \cite{Miller_2004}). 

Since it has been observed that both BH-ULXs and PULXs can become harder when they brighten as mentioned earlier, one cannot rule out the possibility of the presence of these ULX types among the luminous sources in our sample. Therefore, we can conclude that independent of luminosity, harder sources can be possibly NS or massive accretors, but softer and luminous sources with $L_X \geq 10^{40}~erg~s^{-1}$ are most probably massive accretors. However, the type of ULX can only be confirmed with detailed spectral analysis. 

With these assumptions, it can be inferred that many of the ULXs in Figure \ref{fig:lx_ratio_all}, are likely to have neutron stars (sources with HR $>0$) and few may host massive black holes (soft and hard sources with $L_X \geq 10^{40} erg\ s^{-1} $) respectively as accretors. 

Relative distance from the centre could be an indication of the mass of the compact object. A comparison of distances of ULXs in each category from the host centre indicates no obvious dependence of ULX properties on its location in the galaxy. They are randomly distributed, with central sources probably in nuclear star forming regions or galactic bulges. But it should be noted that the most luminous ULXs ie., $L_X \geq 10^{40}erg~s^{-1}$ are mostly located within a relative distance of 0.6 from the centre.

We have examined our sample to look for any known IMBHs and PULX. According to the literature, NGC 2276-3c, ESO 243-49 HLX-1 and M82 X-1 are few of the known IMBH candidates. In our study, NGC 2276-3c belongs to the second group with 3 ULXs, ESO 243-49 HLX-1 being an HLX is not present in the sample and M82 X-1 was flagged
as confused or extended source in all three surveys hence not included in the sample (see \cite{Tranin_2024}). PULXs known so far included in the sample are; M82 X-2, NGC 7793 P13, NGC 5907 ULX1, M51 ULX-7, NGC 300 ULX1 and NGC 1313 X-2. But none of them fall in $N=1$ category except M82 X-2 in $N=1$ luminous spiral category. However, studies indicate the presence of another known ULX (an IMBH M82 X-1) in the same host which is not part of our sample as stated before. 

Using XMM-Newton observations of ULXs, \cite{Earnshaw_2019} concluded that the nature of ULXs in spirals and ellipticals are indistinguishable in hardness ratio space, which is unlike the results we obtain for $N=1$ spirals. The same study also discusses that the majority of ULXs have $L_X < 10^{40} ~erg ~s^{-1}$ and the most luminous ones are hosted by moderately wound spirals ($2\leq$ T $\leq6$).\\

\begin{figure}[hbt!]
\centering
\includegraphics[scale=0.35]{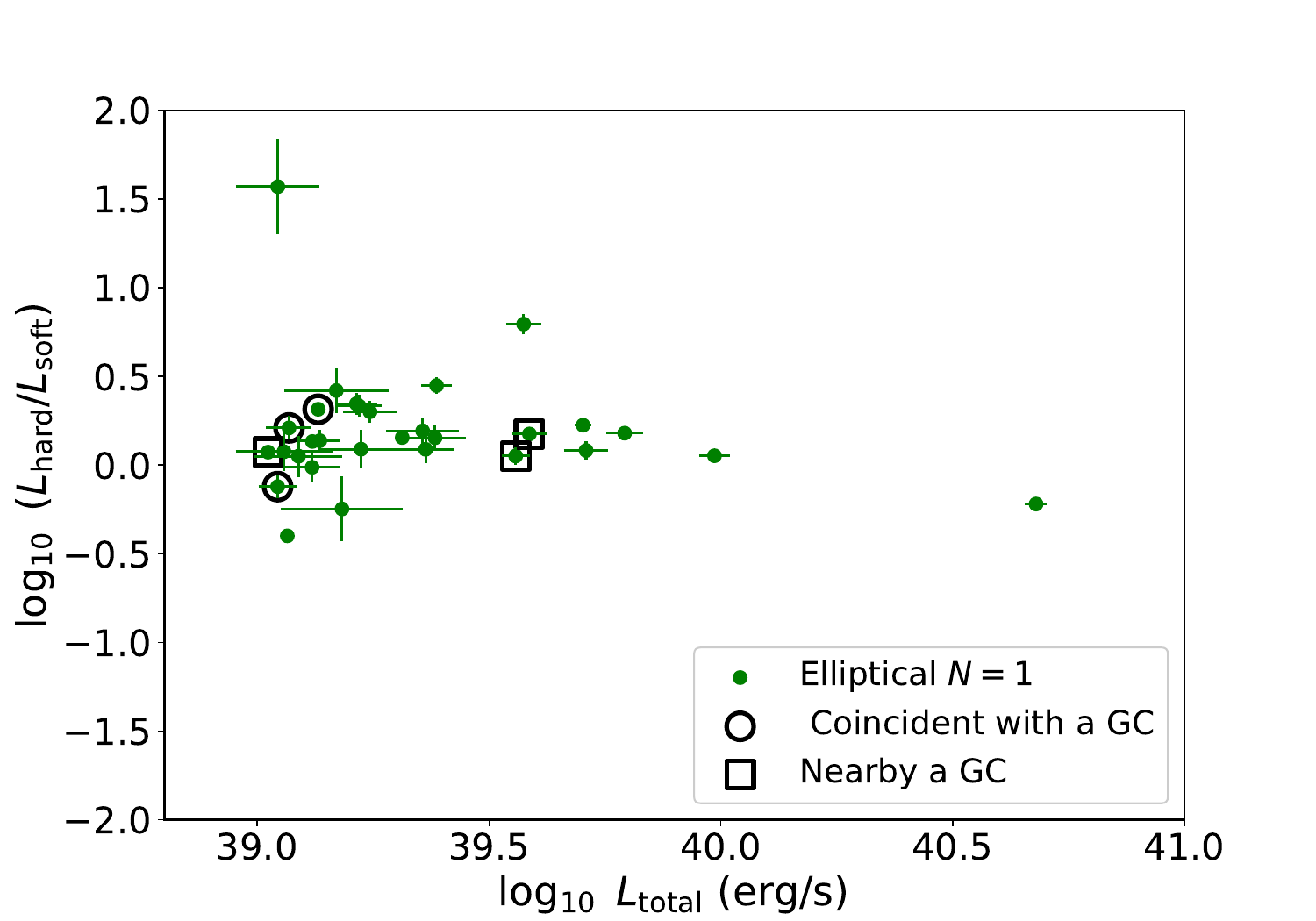}
\caption{log ($L_{hard}$ / $L_{soft}$) vs. broad-band  X-ray luminosity plot for ULXs in $N=1$ ellipticals, showing ULX probably associated with GC and nearby GC marked as black open circles and squares respectively.}
\label{fig:ratio_lx_gc}
\end{figure}

By cross verifying with SIMBAD database \citep{Simbad_2000} we found three ULXs in N=1 elliptical category coincident with GC locations, indicating their origin within the clusters, while three other sources of the same category having a GC nearby ($\geq$ 2 arcsec), indicating they have probably been kicked out from the parent cluster due to stellar interactions (see Fig. \ref{fig:ratio_lx_gc}). It is also noted that, all ULXs in $N\geq7$ elliptical category are located nearby a globular cluster (with nearest 5 arcsec away) which probably shows an association.

\begin{figure}[hbt!]
\centering
\includegraphics[scale=0.35]{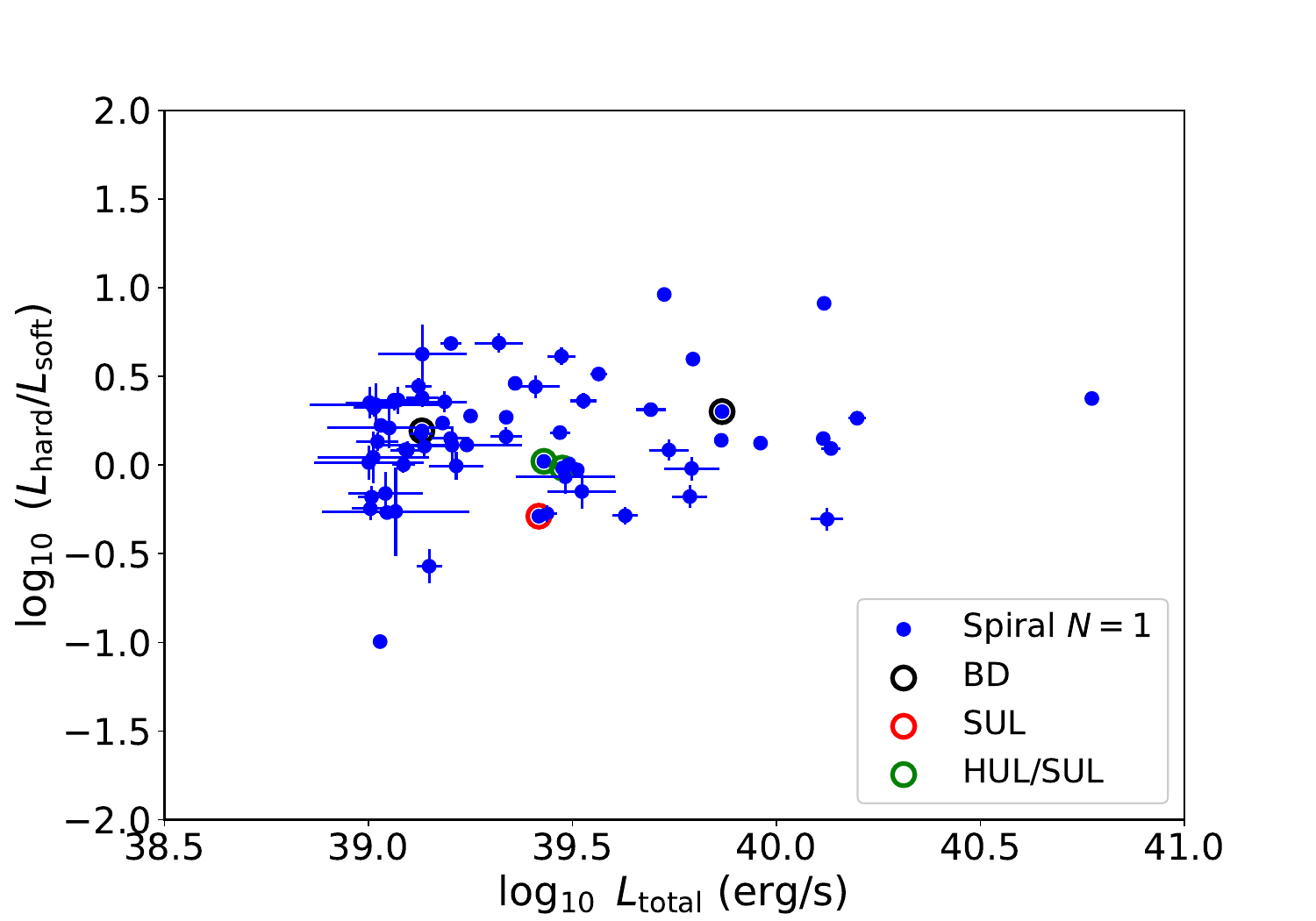}
\caption{log ($L_{hard}$ / $L_{soft}$) vs. broad-band  X-ray luminosity plot for ULXs in $N=1$ spirals with known ULX spectral types in our sample marked as open circles.}
\label{fig:ratio_s_1_types}
\end{figure}

By comparing ULX spectral types described in \cite{Sutton_2013} with our sample, we identified five sources in $N=1$ spiral category (see Fig. \ref{fig:ratio_s_1_types}) : two BD type (NGC 2403 X-1 and NGC 4190 ULX 1), one SUL type (NGC 5408 X-1) and two sources with spectral type changing between SUL and HUL type (NGC 5205 X-1 and NGC 6946 X-1). Among ULXs in \cite{Sutton_2013}, those with $L_X < 3\times10^{39} erg\ s^{-1}$, had BD-type spectra and they suggested that these sources were accreting close to the Eddington rate with $L_X$ limit corresponding to stellar-mass black holes of mass, $M_{BH}$ $\leq$ 20 $M{_\odot}$. Additionally, the high luminosity of BD-type source NGC 4190 ULX 1 with $L_X$ above the BD limit is explained by possibilities such as, over estimation of its distance or having a higher degree of beaming than the fainter disc-like sources or accretion onto massive black holes at Eddington rate \citep{Sutton_2013}. 

\section{Summary and Conclusions}
\label{sect:conclusion}
In this work, a flux limited samples of ULXs occuring at a rate of one per galaxy ($N=1$) and at a larger rate ($N\geq7$) was generated based on the latest ULX catalogue, considering {\it Chandra} and {\it XMM-Newton} observations. We then investigated the dependence of these ULX groups on host properties and a comparative study of source properties among different groups was carried out. We also tried to identify the ULX types in our sample with reference to properties of different categories of ULX sources based on earlier studies. Our major conclusions are as follows; 
\begin{enumerate}
    \item  $N=1$ category ULXs occurs in all spiral types, but for ellipticals they are primarly hosted by star forming type ellipiticals ($-5 \leq$ T $\leq -4$).
    
    \item For the $N=1$ category, ULXs in spirals and ellipticals span a similar range in X-ray luminosity, $L_X$. 
    
    \item Few of the sources have luminosities, $L_X > 10^{40} erg~ s^{-1}$ indicating the possible presence of massive accretors.
    \item Hardness ratios of the $N=1$ spiral category indicates the presence of a reasonable fraction of both soft and hard low luminous sources, and harder high luminous sources. The remaining groups primarily contain harder sources. The K-S test at 95\% confidence level also shows the $N=1$ ULX population in spirals to be different from sources in $N\geq7$ categories.
    \item Six of N=1 elliptical ULXs are associated with GCs (3 coincident with GC location and 3 atleast 2 arcsec from a GC). All ULXs in $N\geq7$ category (hosted by the same galaxy) are atleast 5 arcsec away from a globular cluster.
    \item We find no dependence of ULX properties on their location in the host galaxy, except that the most luminous sources are found relatively closer to the centre. These sources could be massive accretors that have sunk close to the host centre.
\end{enumerate}
In addition, we have identified eight possible massive accretors; six from $N=1$ spiral sample and one each from $N=1$ ellipticals and $N\geq7$ spirals respectively. These very luminous ULXs with $L_{Edd}$ between 1.3 - 5.9 $\times 10^{40}~erg\ s^{-1}$ correspond to black hole masses approximately in the range, $100 -500\ M_\odot$.
The scenerio of massive accretors and their existence is further strengthened by recent gravitational wave observations of events such as GW190521 and GW231123 which produced 142 M$_\odot$ (progenitors of 85 M$_\odot$ and 66 M$_\odot$) and 225 M$_\odot$ black holes (progenitors of 100 M$_\odot$ and 140 M$_\odot$) respectively (see \href{https://www.ligo.caltech.edu/news}{LIGO}\footnote{\url{https://www.ligo.caltech.edu/news}}). This possibility for the existence of massive accretors significantly adds to the importance of studying these luminous sources further.  \\
However, it should be noted that, since ULXs are a diverse and variable class of objects, generalising its features solely based on few available observations could bias the conclusions. Multi-epoch spectral studies are vital in understanding the change in spectral states which can decipher the accretion rate variations. Similarly temporal studies  can probe its flux variability and extract the nature of the accretor. In conclusion, the sample generated as part of this study, will be ideal targets for future multiwavelength, multi-epoch observations that can aid in validating these results and providing deeper insights into the true nature of ULXs. 

\section*{Acknowledgements}
We thank the anonymous referee for careful review of the manuscript. CMS acknowledges the support from Dr. TMA Pai Fellowship of Manipal Academy of Higher Education. This research has made use of data obtained from the Chandra Source Catalog, provided by the Chandra X-ray Center (CXC). This research has made use of data obtained from the 4XMM XMM-Newton serendipitous source catalogue compiled by the 10 institutes of the XMM-Newton Survey Science Centre selected by ESA. This research has made use of the SIMBAD database, operated at CDS, Strasbourg, France.

\onecolumn
\appendix
\section{ULXs in elliptical galaxies}
\label{sec:elliptical_dataset}
\begin{longtable}{l c c c c}
\caption{ULXs in N=1  elliptical category} \\
\hline
 \shortstack{Source No.\\ \ }& \shortstack{RA (J2000)\\(deg)} & \shortstack{DEC (J2000)\\(deg)}& \shortstack{$L_X$ \\($10^{39} erg s^{-1}$)} &  \shortstack{log\ ($L_{\rm hard}/L_{\rm soft}$)\\ \  }\\
\hline
\endfirsthead

\multicolumn{4}{c}{Table A.8 -- continuation} \\
\hline
RA (J2000) & DEC (J2000) & $L_X$ ($10^{39}$ erg s$^{-1}$) & $L_{\rm hard}/L_{\rm soft}$ \\
& (deg)&(deg)& ($10^{39} erg\ s^{-1}$) & \\
\hline
\endhead

\hline
\multicolumn{4}{r}{Continued on next page} \\
\endfoot

\hline
\endlastfoot
1 & 198.23169 & -19.51119 & 5.04 $\pm$ 0.21 & 0.225 $\pm$ 0.019\\
2 & 356.08368 & 9.93011 & 2.42 $\pm$ 0.37  & 0.1540 $\pm$ 0.070\\
3 & 185.11546 & 75.35688 & 2.31 $\pm$ 0.32  & 0.089 $\pm$ 0.076\\
4 & 186.34518 & 18.18149 & 1.11 $\pm$ 0.10 & -0.121 $\pm$ 0.066\\
5 & 306.9144 & -47.02681 & 1.14 $\pm$ 0.27  & 0.075 $\pm$ 0.108\\
6 & 208.36874 & 40.31194 & 1.67 $\pm$ 0.34 & 0.089 $\pm$ 0.109\\
7 & 207.2502 & 60.20186 & 1.48 $\pm$ 0.38  & 0.420 $\pm$ 0.127\\
8 & 133.91824 & 58.71806 & 1.75 $\pm$ 0.23  & 0.301 $\pm$ 0.063\\
9 & 181.16045 & 1.78782 & 47.9 $\pm$ 0.26  & -0.219 $\pm$ 0.028\\
10 & 224.84064 & -16.63108 & 2.27 $\pm$ 0.41  & 0.1930 $\pm$ 0.073\\
11 & 178.68472 & -13.98178 & 1.52 $\pm$ 0.46  & -0.247 $\pm$ 0.182\\
12 & 187.26418 & 13.97119 & 1.05 $\pm$ 0.05 & 0.073 $\pm$ 0.024\\
13 & 157.6297 & -35.36665 & 6.20 $\pm$ 0.57 & 0.180 $\pm$ 0.037\\
14 & 316.36609 & -52.57265 & 1.31 $\pm$ 0.18 & -0.012 $\pm$ 0.079\\
15 & 229.64326 & -24.07852 & 1.66 $\pm$ 0.18 & 0.333 $\pm$ 0.062\\
16 & 201.42615 & -42.99545 & 1.16 $\pm$ 0.02 & -0.399 $\pm$ 0.014\\
17 & 186.91951 & 13.07992 & 3.86 $\pm$ 0.33  & 0.175 $\pm$ 0.030\\
18 & 114.56513 & -69.49108 & 1.11 $\pm$ 0.23  & 1.570 $\pm$ 0.268\\
19 & 169.62071 & 57.99566 & 3.75 $\pm$ 0.33  & 0.795 $\pm$ 0.059\\
20 & 157.2897 & -35.59883 & 9.69 $\pm$ 0.74  & 0.053 $\pm$ 0.036\\
21 & 157.2435 & -35.59703 & 1.23 $\pm$ 0.27  & 0.050 $\pm$ 0.116\\
22 & 210.91074 & -33.96507 & 5.12 $\pm$ 0.57  & 0.082 $\pm$ 0.051\\
23 & 33.51701 & 27.87763 & 1.31 $\pm$ 0.05  & 0.133 $\pm$ 0.021\\
24 & 55.57692 & -35.39395 & 1.17 $\pm$ 0.13  & 0.210 $\pm$ 0.066\\
25 & 161.95831 & 12.58246 & 2.05 $\pm$ 0.03  & 0.155 $\pm$ 0.008\\
26 & 302.47888 & -48.40023 & 2.44 $\pm$ 0.18  & 0.448 $\pm$ 0.045\\
27 & 344.28967 & -36.47339 & 1.64 $\pm$ 0.17  & 0.346 $\pm$ 0.060\\
28 & 54.30669 & -35.73033 & 3.61 $\pm$ 0.25  & 0.051 $\pm$ 0.049\\
29 & 188.92175 & 12.58092 & 1.35 $\pm$ 0.05  & 0.315 $\pm$ 0.019\\
30 & 187.869	 & 25.74939 & 1.36 $\pm$ 0.13 & 0.137 $\pm$ 0.059\\
\end{longtable}

\begin{longtable}{l c c c c}
\caption{ULXs in $N\geq7$  elliptical category} \\
\hline
 \shortstack{Source No.\\ \ }& \shortstack{RA (J2000)\\(deg)} & \shortstack{DEC (J2000)\\(deg)}& \shortstack{$L_X$ \\($10^{39} erg s^{-1}$)} &  \shortstack{log\ ($L_{\rm hard}/L_{\rm soft}$)\\ \  }\\
\hline
\endfirsthead

\multicolumn{4}{c}{Table A.9 -- continuation} \\
\hline
RA (J2000) & DEC (J2000) & $L_X$ ($10^{39}$ erg s$^{-1}$) & $L_{\rm hard}/L_{\rm soft}$ \\
& (deg)&(deg)& ($10^{39} erg\ s^{-1}$) & \\
\hline
\endhead

\hline
\multicolumn{4}{r}{Continued on next page} \\
\endfoot

\hline
\endlastfoot

1 & 55.03715 & -18.57994&1.08$\pm$0.11&-0.139$\pm$0.081\\
2 & 55.03856 & -18.57815&1.1$\pm$0.14&0.314$\pm$0.077\\
3 & 55.04424 & -18.59531&1.19$\pm$0.15&0.325$\pm$0.07\\
4 & 55.04677 & -18.5764&1.98$\pm$0.2&-0.042$\pm$0.071\\
5 & 55.05216 & -18.581&5.41$\pm$0.31&0.042$\pm$0.045\\
6 & 55.06051 & -18.61052&2.12$\pm$0.17&0.307$\pm$0.043\\

\end{longtable}

\onecolumn
\clearpage

\section{ULXs in spiral galaxies}
\label{spiral_dataset}
\begin{longtable}{l c c c c}
\caption{ULX in N=1 spiral category} \\
\hline
 \shortstack{Source No.\\ \ }& \shortstack{RA (J2000)\\(deg)} & \shortstack{DEC (J2000)\\(deg)}& \shortstack{$L_X$ \\($10^{39} erg s^{-1}$)} &  \shortstack{log\ ($L_{\rm hard}/L_{\rm soft}$)\\ \  }\\
\hline
\endfirsthead

\multicolumn{4}{c}{Table B.10 -- continuation} \\
\hline
 \shortstack{Source No.\\ \ }& \shortstack{RA (J2000)\\(deg)} & \shortstack{DEC (J2000)\\(deg)}& \shortstack{$L_X$ \\($10^{39} erg s^{-1}$)} &  \shortstack{log\ ($L_{\rm hard}/L_{\rm soft}$)\\ \  }\\
\hline
\endhead

\hline
\multicolumn{4}{r}{Continued on next page} \\
\endfoot

\hline
\endlastfoot
1 & 198.38081&-19.54377&4.92 $\pm$ 0.41&0.313 $\pm$ 0.03\\
2 & 200.57645&-16.71301&59.31 $\pm$ 1.52&0.375 $\pm$ 0.012\\
3 & 148.96011&69.67943&13.09 $\pm$ 0.05&0.912 $\pm$ 0.003\\
4 & 180.47852&-18.88689&9.14 $\pm$ 0.29&0.124 $\pm$ 0.019\\
5 & 13.77028&-37.69553&1.78 $\pm$ 0.01&0.277 $\pm$ 0.002\\
6 & 354.09711&2.15762&1.03 $\pm$ 0.32&0.044 $\pm$ 0.147\\
7 & 354.77856&-12.22515&1.12 $\pm$ 0.39&0.21 $\pm$ 0.115\\
8 & 248.10286&78.18657&1.64 $\pm$ 0.25&-0.005 $\pm$ 0.078\\
9 & 349.0675&-42.58823&2.74 $\pm$ 0.16&-0.274 $\pm$ 0.048\\
10 & 14.87362&-36.18606&3.04 $\pm$ 0.85&-0.066 $\pm$ 0.096\\
11 & 186.41301&12.66914&1.59 $\pm$ 0.17&0.15 $\pm$ 0.034\\
12 & 60.07106&-67.80436&1.01 $\pm$ 0.10&-0.246 $\pm$ 0.067\\
13 & 354.06522&2.15654&13.61 $\pm$ 0.74&0.094 $\pm$ 0.03\\
14 & 186.50597&33.52534&1.07 $\pm$ 0.01&-0.995 $\pm$ 0.027\\
15 & 154.40569&21.69569&1.37 $\pm$ 0.17&0.107 $\pm$ 0.069\\
16 & 213.88631&36.22945&1.17 $\pm$ 0.48&-0.261 $\pm$ 0.25\\
17 & 223.47762&3.56212&1.18 $\pm$ 0.18&0.367 $\pm$ 0.076\\
18 & 340.22644&75.15672&1.04 $\pm$ 0.39&0.341 $\pm$ 0.122\\
19 & 185.61247&29.90159&1.03 $\pm$ 0.12&0.324 $\pm$ 0.053\\
20 & 188.17793&0.11528&6.24 $\pm$ 0.06&0.598 $\pm$ 0.006\\
21 & 107.45714&-27.57662&6.14 $\pm$ 0.60&-0.177 $\pm$ 0.066\\
22 & 53.40928&-5.09718&5.45 $\pm$ 0.62&0.085 $\pm$ 0.059\\
23 & 157.35255&29.51046&1.54 $\pm$ 0.19&0.357 $\pm$ 0.058\\
24 & 185.40019&14.59574&1.11 $\pm$ 0.05&-0.268 $\pm$ 0.034\\
25 & 248.81366&-58.09948&2.57 $\pm$ 0.35&0.442 $\pm$ 0.065\\
26 & 147.53206&-73.92354&1.24 $\pm$ 0.12&0.084 $\pm$ 0.049\\
27 & 183.76567&14.03044&1.60 $\pm$ 0.63&0.112 $\pm$ 0.107\\
28 & 324.13287&-54.56593&2.29 $\pm$ 0.08&0.461 $\pm$ 0.017\\
29 & 170.05264&67.24503&1.00 $\pm$ 0.31&0.015 $\pm$ 0.096\\
30 & 353.32156&-54.08346&6.20 $\pm$ 0.97&-0.019 $\pm$ 0.066\\
31 & 109.15948&-62.33695&15.77 $\pm$ 0.78&0.264 $\pm$ 0.024\\
32 & 147.49059&-25.00523&13.31 $\pm$ 1.21&-0.305 $\pm$ 0.065\\
33 & 143.02583&21.51637&1.07 $\pm$ 0.02&0.224 $\pm$ 0.012\\
34 & 185.80521&11.35707&1.05 $\pm$ 0.12&0.132 $\pm$ 0.052\\
35 & 188.62598&9.62505&1.16 $\pm$ 0.16&0.356 $\pm$ 0.047\\
36 & 31.14315&-6.20263&1.41 $\pm$ 0.10&-0.57 $\pm$ 0.096\\
37 & 293.32452&-58.12179&1.10 $\pm$ 0.23&-0.161 $\pm$ 0.121\\
38 & 344.15494&-37.04154&2.09 $\pm$ 0.29&0.688 $\pm$ 0.054\\
39 & 185.31709&11.51905&13.02 $\pm$ 0.26&0.149 $\pm$ 0.009\\
40 & 194.05258&-29.50067&3.33 $\pm$ 0.64&-0.149 $\pm$ 0.095\\
41 & 156.97981&-43.91302&7.33 $\pm$ 0.238&0.14 $\pm$ 0.017\\
42 & 202.41093&58.41821&2.69 $\pm$ 0.03&0.021 $\pm$ 0.006\\
43 & 21.13953&3.79658&1.35 $\pm$ 0.34&0.626 $\pm$ 0.165\\
44 & 146.46034&-14.36048&4.26 $\pm$ 0.30&-0.285 $\pm$ 0.05\\
45 & 169.74368&13.09176&2.18 $\pm$ 0.08&0.27 $\pm$ 0.018\\
46 & 183.43867&36.63189&7.36 $\pm$ 0.05&0.302 $\pm$ 0.004\\
47 & 16.29703&-6.20196&2.17 $\pm$ 0.19&0.16 $\pm$ 0.053\\
48 & 40.66204&-0.01529&5.31 $\pm$ 0.15&0.962 $\pm$ 0.021\\
49 & 146.37199&-31.2077&1.16 $\pm$ 0.05&0.366 $\pm$ 0.031\\
50 & 154.99664&45.56758&1.23 $\pm$ 0.06&0.087 $\pm$ 0.034\\
51 & 165.85727&18.14552&2.95 $\pm$ 0.16&0.183 $\pm$ 0.034\\
52 & 214.91414&56.6938&3.36 $\pm$ 0.24&0.362 $\pm$ 0.043\\
53 & 171.87476&56.88711&1.35 $\pm$ 0.12&0.38 $\pm$ 0.053\\
54 & 177.29971&56.08601&1.02 $\pm$ 0.08&-0.18 $\pm$ 0.064\\
55 & 181.38472&50.54605&2.97 $\pm$ 0.24&0.614 $\pm$ 0.05\\
56 & 183.18544&10.85379&3.67 $\pm$ 0.17&0.514 $\pm$ 0.03\\
57 & 183.95456&47.08895&1.22 $\pm$ 0.08&0.002 $\pm$ 0.044\\
58 & 188.45067&15.15491&1.33 $\pm$ 0.1&0.444 $\pm$ 0.046\\
59 & 227.45667&57.00006&1.74 $\pm$ 0.08&0.114 $\pm$ 0.03\\
60 & 204.37553&8.87595&1.35 $\pm$ 0.09&0.176 $\pm$ 0.041\\
61 & 40.10684&-8.40846&1.52 $\pm$ 0.03&0.237 $\pm$ 0.013\\
62 & 94.80838&78.36251&1.59 $\pm$ 0.09&0.686 $\pm$ 0.039\\
63 & 114.10648&65.59441&1.35 $\pm$ 0.01&0.193 $\pm$ 0.005\\
64 & 194.75775&34.85408&3.25 $\pm$ 0.13&-0.027 $\pm$ 0.03\\
65 & 210.83176&-41.38297&2.62 $\pm$ 0.02&-0.288 $\pm$ 0.008\\
66 & 222.3862&-10.16324&3.10 $\pm$ 0.14&0.008 $\pm$ 0.03\\
67 & 308.75308&60.19188&2.99 $\pm$ 0.02&-0.016 $\pm$ 0.005\\
68 & 349.73303&-42.23334&1.01 $\pm$ 0.14&0.352 $\pm$ 0.089\\
\end{longtable}

\begin{longtable}{l c c c c}
\caption{ULXs in $N\geq7$  spiral category} \\
\hline
 \shortstack{Source No.\\ \ }& \shortstack{RA (J2000)\\(deg)} & \shortstack{DEC (J2000)\\(deg)}& \shortstack{$L_X$ \\($10^{39} erg s^{-1}$)} &  \shortstack{log\ ($L_{\rm hard}/L_{\rm soft}$)\\ \  }\\
\hline
\endfirsthead

\multicolumn{4}{c}{Table A.9 -- continuation} \\
\hline
 \shortstack{Source No.\\ \ }& \shortstack{RA (J2000)\\(deg)} & \shortstack{DEC (J2000)\\(deg)}& \shortstack{$L_X$ \\($10^{39} erg s^{-1}$)} &  \shortstack{$L_{\rm hard}/L_{\rm soft}$\\ \  }\\
\hline
\endhead

\hline
\multicolumn{4}{r}{Continued on next page} \\
\endfoot

\hline
\endlastfoot

1 & 85.5055&69.37759&15.92$\pm$2.07&0.797$\pm$0.099\\
2  & 85.55151&69.36555&7.43$\pm$1.37&0.698$\pm$0.119\\
3  & 94.06628&-21.37574&4.35$\pm$0.29&0.193$\pm$0.044\\
4  & 94.06943&-21.37434&1.27$\pm$0.16&0.139$\pm$0.078\\
5  & 94.07054&-21.35273&2.61$\pm$0.28&0.218$\pm$0.066\\
6  & 94.07186&-21.38069&3.68$\pm$0.3&0.367$\pm$0.05\\
7  & 94.07511&-21.36474&1.19$\pm$0.17&0.362$\pm$0.086\\
8  & 94.07828&-21.37434&2.24$\pm$0.22&0.326$\pm$0.062\\
9  & 94.08686&-21.38082&1.34$\pm$0.19&0.282$\pm$0.082\\
10  & 94.09799&-21.37188&1.3$\pm$0.21&0.451$\pm$0.09\\
11 & 94.10491&-21.37487&1.17$\pm$0.18&0.409$\pm$0.1\\
12 & 94.11056&-21.37015&1.9$\pm$0.2&0.144$\pm$0.066\\
13 & 94.11184&-21.36983&1.11$\pm$0.13&0.091$\pm$0.084\\
14 & 94.11826&-21.3609&1.11$\pm$0.16&0.126$\pm$0.086\\
15 & 161.04059&6.75573&6.72$\pm$1.17&0.813$\pm$0.123\\
16 & 161.04587&6.76109&9.78$\pm$1.19&0.226$\pm$0.071\\
17 & 304.23663&-70.77689&6.32$\pm$0.78&0.388$\pm$0.077\\
18 & 304.25375&-70.7692&5.85$\pm$1.02&1.252$\pm$0.219\\
19 & 304.24084&-70.78432&7.31$\pm$1.01&0.629$\pm$0.084\\
20 & 94.075&-21.3678&5.7$\pm$0.33&0.26$\pm$0.034\\
\end{longtable}

\twocolumn






\end{document}